\documentclass[journal,10pt]{IEEEtran}
\usepackage{amsmath,amssymb,amsfonts}
\usepackage{graphicx}
\usepackage{float}
\usepackage{subfig}
\usepackage{color}
\usepackage{cite}
\usepackage{multirow}
\usepackage{algorithm}  
\usepackage{algorithmicx}  
\usepackage{algpseudocode} 
\usepackage{bm}
\usepackage[colorlinks,linkcolor=black,anchorcolor=black,
citecolor=black,urlcolor=black]{hyperref}

\usepackage{stfloats}
\usepackage{siunitx}
\DeclareSIUnit\bps{bps}
\DeclareSIUnit\Torr{Torr}
\DeclareSIUnit\torr{Torr}
\DeclareSIUnit\sample{Sa}
\newcommand*{\circled}[1]{\lower.7ex\hbox{\tikz\draw (0pt, 0pt)%
  circle (.5em) node {\makebox[1em][c]{\small #1}};}}

\ifCLASSINFOpdf

\else

\fi

\graphicspath{{figures/}}

\begin{document}

\title{DTECM: Digital Twin Enabled Channel Measurement and Modeling in Terahertz Urban Macrocell}
\author{Yuanbo~Li, Ziming~Yu, Chong~Han,~\IEEEmembership{Senior~Member,~IEEE}
\thanks{
This paper was presented in part at the IEEE PIMRC, 2024~\cite{Li2024220}. 

Yuanbo Li and Chong Han are with the Terahertz Wireless Communications (TWC) Laboratory, Shanghai Jiao Tong University, Shanghai, China (e-mail: \{yuanbo.li, chong.han\}@sjtu.edu.cn).

Ziming Yu is with Huawei Technologies Co., Ltd, China (e-mail:yuziming@huawei.com).
}
}

\maketitle

\begin{abstract}
To achieve ultra-fast communication speeds in future communication networks, the Terahertz band, ranging from 0.1~THz to 10~THz, is envisioned as a key technology due to its abundant spectrum resources and broad consecutive usable bandwidth. The realization of reliable THz communications demands a full understanding and accurate characterization of THz channels. As various channel measurements and modeling studies have been conducted recently, the urban macrocell (UMa) scenario is still not fully investigated considering the large cell radius and foliage losses. In this work, in the THz UMa, extensive channel measurements are conducted and an accurate channel model is developed by combining ray-tracing, computer vision (CV), and statistical methods. Specifically, substantial channel measurement campaigns with distances up to 410~m are conducted at 220~GHz, with nanosecond-level absolute time synchronization. Based on the measurement results, the propagation phenomena are analyzed in detail and the channel characteristics are calculated and statistically modeled. Furthermore, a digital twin enabled channel model (DTECM) is proposed, which generates THz channel responses in a hybrid manner. Specifically, the dominant paths are generated deterministically by using the ray-tracing technique and CV methods. Apart from the path gains determined by ray-tracing, the additional foliage loss is accurately modeled based on foliage information extracted from panoramic pictures. To maintain a low computational complexity for the DTECM, non-dominant paths are then generated statistically. Numeric results reveal that compared to the traditional statistical channel models, the DTECM reduces the path loss modeling error from 14~dB to 4~dB, showing its great superiority. Furthermore, a preliminary link performance evaluation using the DTECM indicates that THz UMa is feasible, though requiring high antenna gains and coverage extension techniques to achieve high spectral efficiencies and wide coverage.
\end{abstract}

\IEEEpeerreviewmaketitle

\section{Introduction}
\par \IEEEPARstart{O}{ver} the past several decades, mobile communication technologies have undergone rapid evolution, progressing from first-generation (1G) systems to the current fifth-generation (5G) networks. Key advancements in 5G, such as massive multiple-input multiple-output (MIMO), millimeter-wave communications, network virtualization, and softwarization, have delivered substantial benefits across commercial, industrial, and consumer sectors~\cite{Akyildiz20206g}. However, emerging applications including holographic teleportation, the metaverse, and autonomous driving present requirements that exceed the capabilities of 5G, thereby motivating the development of next-generation communication systems~\cite{Saad2020vision,Zhang20196G}. 
To support the exponential growth in data traffic driven by intelligent applications and interconnected devices, future networks are expected to achieve data rates on the order of hundreds of gigabits per second, potentially reaching the terabit-per-second (Tbps) level~\cite{chen2021terahertz}. In this context, the Terahertz (THz) band, spanning 0.1–10~THz, is envisioned as a key enabler of ultra-high-speed communications~\cite{akyildiz2018combating,rappaport2019wireless}. Owing to its abundant spectrum resources and extremely wide bandwidth (exceeding tens of gigahertz), the THz band offers a promising solution to overcome the spectral congestion and capacity limitations of existing wireless systems.
\par A fundamental challenge in realizing THz wireless communications is the comprehensive understanding of THz wave propagation mechanisms and the accurate characterization of THz channels. Among various methods, channel measurement campaigns remain the most credible and widely adopted approach for obtaining realistic channel data. Over the past decade, extensive measurement efforts have been undertaken across diverse indoor and outdoor scenarios, such as offices, conference rooms, corridors, data centers, street canyons, and urban microcells (UMi), at representative frequencies including \SI{140}{GHz}, \SI{220}{GHz}, and \SI{300}{GHz}~\cite{han2022terahertz}. Notably, the research group at Technische Universität Braunschweig has conducted channel measurements since 2011, focusing on environments such as conference rooms, offices, and vehicle-to-vehicle scenarios primarily around 300 GHz~\cite{priebe2011channel,priebe2013ultra,Eckhardt2021channel}. The University of Southern California has investigated outdoor urban channels near \SI{140}{GHz} and \SI{220}{GHz}~\cite{abbasi2020double,abbasi2021double,Abbasi2023THz}, while New York University has studied factory, office, and UMi environments at \SI{140}{GHz}~\cite{Xing2021Indoor,Xing2021millimeter,Shakya2024Radio,Ju2024142}. Similarly, the Beijing University of Posts and Telecommunications has conducted measurements in office, factory, and UMi scenarios at \SI{105}{GHz} and \SI{132}{GHz}~\cite{Qin2024Time,Liu2024Channel,Chang20233GPP}. In our recent work, we have also performed comprehensive measurement campaigns in both indoor and outdoor environments across a broad frequency range from \SI{140}{GHz} to \SI{400}{GHz}~\cite{chen2021channel,wang2023thz,Li2023Correlation,li2024pico,Wang2025300}.
Despite these efforts, most outdoor channel measurements have been limited to relatively short distances (typically around \SI{100}{m}). The urban macrocell (UMa) scenario, characterized by larger cell radii on the order of \SIrange{400}{500}{m}, remains insufficiently explored. This paper aims to address this gap by conducting extensive THz channel measurements and modeling in the UMa environment.
\par The ultimate objective of channel studies is to develop accurate and efficient channel models that serve as standardized baselines for communication system design and performance evaluation. For the THz band, the research community has explored a range of modeling approaches, including deterministic methods~\cite{Kanhere2025Calibration,Yi2023Full,Azpilicueta2023Diffuse}, statistical methods~\cite{Ju2024Statistical,Poddar2024Tutorial,hu2024Transfer,Wang2023Novel}, and hybrid methods~\cite{chen2021channel,Wang2025300,Eckhardt2024Hybrid,Takahashi2025Channel}. Deterministic approaches, such as ray-tracing simulations, offer high accuracy but incur significant computational complexity. In contrast, statistical models are computationally efficient but generally less accurate. To leverage the strengths of both approaches, recent efforts have focused on hybrid models that aim to achieve a balance between accuracy and complexity.
Despite these advancements, a critical challenge that remains unresolved in existing THz channel models is the accurate modeling of foliage-induced attenuation in outdoor environments. Statistical models, which lack site-specific information, fail to provide sufficient accuracy in capturing foliage loss. Meanwhile, deterministic models depend on detailed environmental data, and modeling foliage with high fidelity is infeasible due to the vast number and variability of leaves. Therefore, novel modeling approaches are required to effectively characterize foliage loss and support accurate THz channel modeling in UMa scenarios.
\par In this paper, we present extensive channel measurements in the THz UMa scenario at \SI{220}{GHz} using a correlation-based channel sounder. A total of 72 Rx locations are measured, with link distances ranging from \SI{34}{m} to \SI{410}{m}. Based on the measurement results, detailed propagation analysis is conducted, and key channel characteristics are extracted. To accurately model the THz UMa channel, we propose a hybrid channel model, termed the digital twin enabled channel model (DTECM), which integrates ray-tracing, computer vision (CV) techniques, and statistical methods. In particular, foliage-induced attenuation is addressed by constructing a foliage digital twin from panoramic images in the angular domain, enabling accurate foliage loss modeling. Comparative results show that the DTECM significantly outperforms traditional statistical models in modeling accuracy and effectiveness for the THz UMa scenario. Finally, link-level performance evaluations using the DTECM demonstrate that achieving reliable THz connectivity in UMa environments requires high antenna gains and coverage extension techniques. The main contributions of this work are summarized as follows.
\begin{itemize}
    \item \textbf{We conduct extensive channel measurements in the THz UMa at 220~GHz using a correlation-based channel sounder.} A total of 72 Rx locations are measured, encompassing line-of-sight (LoS), obstructed-LoS (OLoS), and non-LoS (NLoS) cases, with distances between the transmitter (Tx) and Rx ranging from \SI{34}{m} to \SI{410}{m}. To the best of the authors' knowledge, this is the first work to report long-distance THz channel measurements exceeding \SI{400}{m}, made possible by the high performance up/down converters, significant baseband processing gain, and nanosecond-level absolute time synchronization.
    \item \textbf{We perform detailed propagation analysis and channel characterization in the THz UMa.} Based on the measured power-delay profiles (PDPs), it is observed that the foliage blockage significantly affects the strength of the LoS path, while some typical objects, such as buildings, cars, guideboards, and metal pillars, etc., incur weak reflection and scattering paths. Additionally, channel characteristics are calculated and thoroughly analyzed.
    \item \textbf{We model the additional loss due to foliage blockage using panoramic images and CV methods.} A k-nearest neighbor (KNN)-based foliage identification approach is employed to extract foliage from panoramic images, with camera pose correction applied to accurately determine the angular positions of the foliage. Using this foliage data, the additional loss from foliage blockage is modeled as a linear function of the foliage coverage ratio (FCR).
    \item \textbf{We propose the DTECM, which integrates ray-tracing, CV methods, and statistical models for accurate channel modeling in the THz UMa.} The dominant multipath components (MPCs) in the THz channels, including LoS paths and once-reflected paths, are generated deterministically using ray-tracing, with path gains further adjusted based on foliage information extracted from panoramic images. The remaining weak MPCs are generated statistically. Numerical results demonstrate that the proposed DTECM outperforms traditional statistical models in terms of accuracy. Additionally, link-level performance in the THz UMa is evaluated using the DTECM, with results highlighting the critical role of high antenna gains and coverage extension techniques in establishing effective links in the THz UMa.
\end{itemize}
\par The remainder of the paper is organized as follows. The channel measurement campaigns are introduced in Sec.~\ref{sec:measurement}. Section~\ref{sec:char} provides detailed propagation analysis and channel characterization in the THz UMa. Furthermore, in Sec.~\ref{sec:dtecm}, the DTECM is proposed, validated, and discussed. Finally, Sec.~\ref{sec:conclude} concludes the paper.
\section{Channel Measurement Campaign}
\label{sec:measurement}
\subsection{Measurement System}
\label{sec:system}
\begin{figure}
    \centering
    \includegraphics[width=0.9\columnwidth]{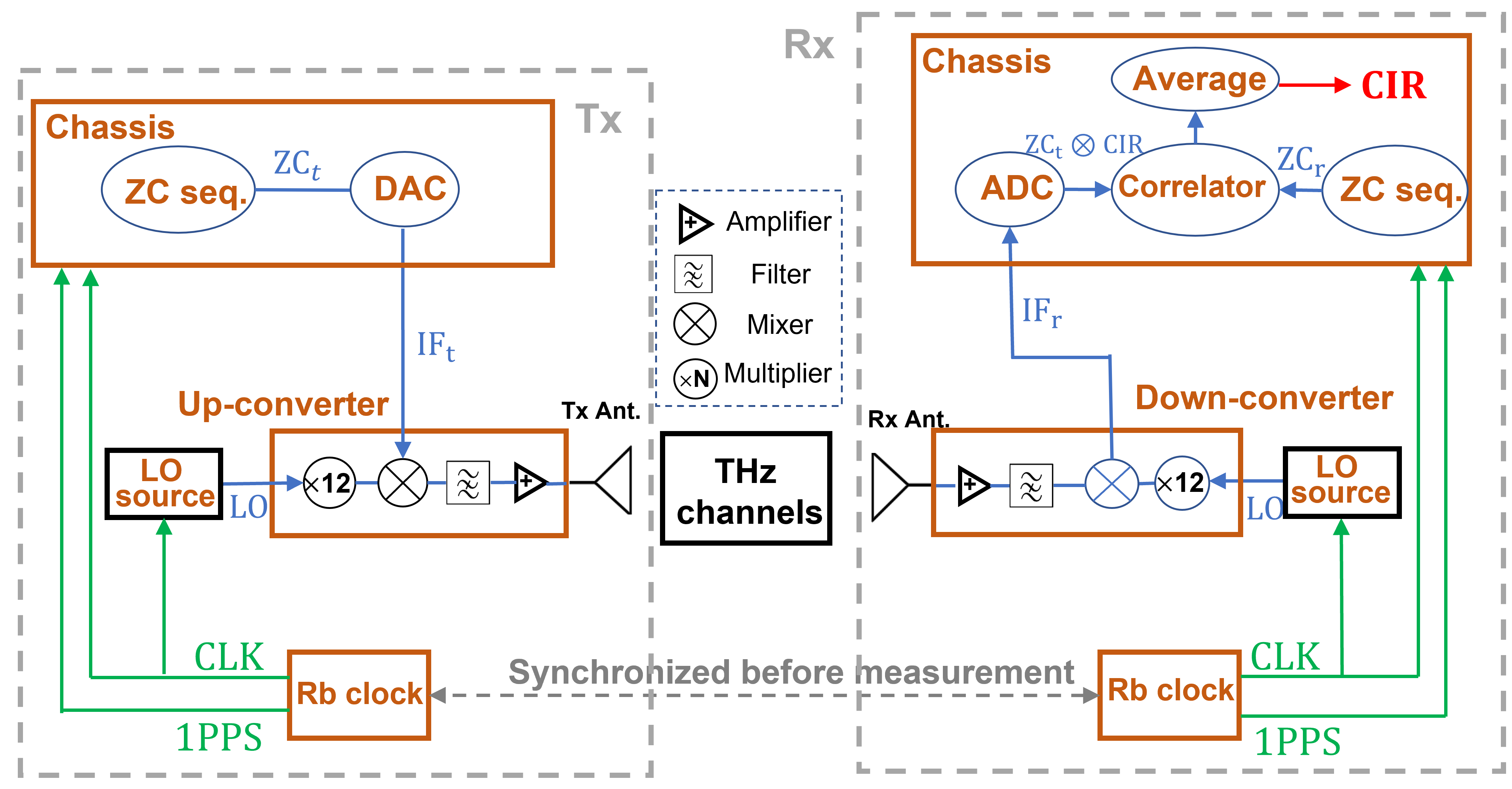}  
    \caption{Correlation-based time-domain channel sounder.}
    \label{fig:cs}
    \vspace{-0.5cm}
\end{figure}
\par To measure THz channels, a correlation-based time-domain channel sounder is used, as shown in Fig.~\ref{fig:cs}. Both the Tx and Rx include a baseband processing chassis, a radio frequency front end, and a mechanical platform for the movement of Tx/Rx. The channel impulse response is measured through the auto-correlation of a Zadoff-Chu (ZC) sequence. Additionally, two rubidium (Rb) clocks are utilized to separately provide clock reference and 1 pulse-per-second (1PPS) triggers to the Tx and Rx subsystems, which are connected in a master-slave mode for several hours before real measurements to synchronize the 1PPS signal. For a detailed description, readers may refer to our previous work in\cite{Li2023Correlation}.
\subsection{Measurement Campaign}
\par The measurement campaign is conducted near Nanyangdong Road on the Shanghai Jiao Tong University campus. Buildings with heights ranging from \SI{10}{m} to \SI{20}{m} are distributed across the campus, including the Longbin Building, Wenbo Building, and buildings of the School of Mechanical Engineering (ME), among others. Additionally, evergreen trees, with heights of approximately \SI{5}{m}, are densely planted throughout the campus.
\begin{figure*}[!tbp]
    \centering
    \includegraphics[width=1.75\columnwidth]{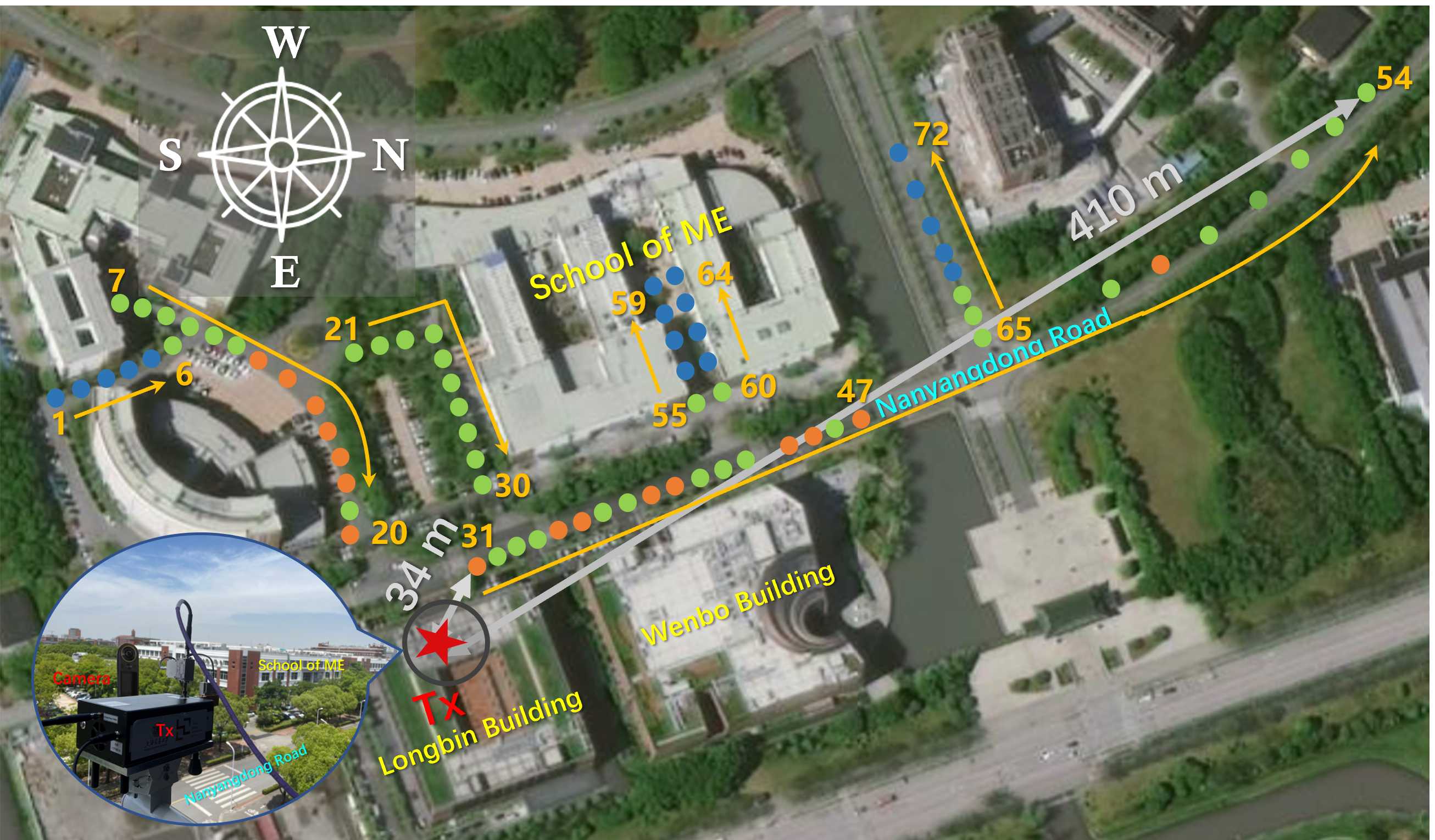}  
    \caption{The measurement deployment in the UMa scenario. Rx positions are labeled with colored dots. Orange dots represent Rx positions with LoS propagation, while green dots and blue dots are those in OLoS and NLoS conditions.}
    \label{fig:layout}
    \vspace{-0.5cm}
\end{figure*}
\par To simulate a UMa base station, the Tx is deployed on a horizontal pole extending out of a window on the fourth floor of the Longbin Building, achieving a height of \SI{16.6}{m} above the ground, as shown in the inset picture in Fig.~\ref{fig:layout}. A fisheye camera is mounted on the Tx to capture panoramic images of the measured scenario, which are later used for digital twin generation. The Rx, representing user equipment, is positioned at a height of \SI{1.6}{m}, with locations indicated by colored dots in Fig.~\ref{fig:layout}. In total, 72 Rx positions are measured, including 19 with clear LoS propagation, 33 partially obstructed by foliage or buildings, and 20 in NLoS areas. The Tx-Rx separation distance ranges from \SI{34}{m} to \SI{410}{m}.
\subsection{Measurement Setup and Link Budget Analysis}







        
        
        






\begin{figure*}[!tbp]
    \centering
    \includegraphics[width=1.8\columnwidth]{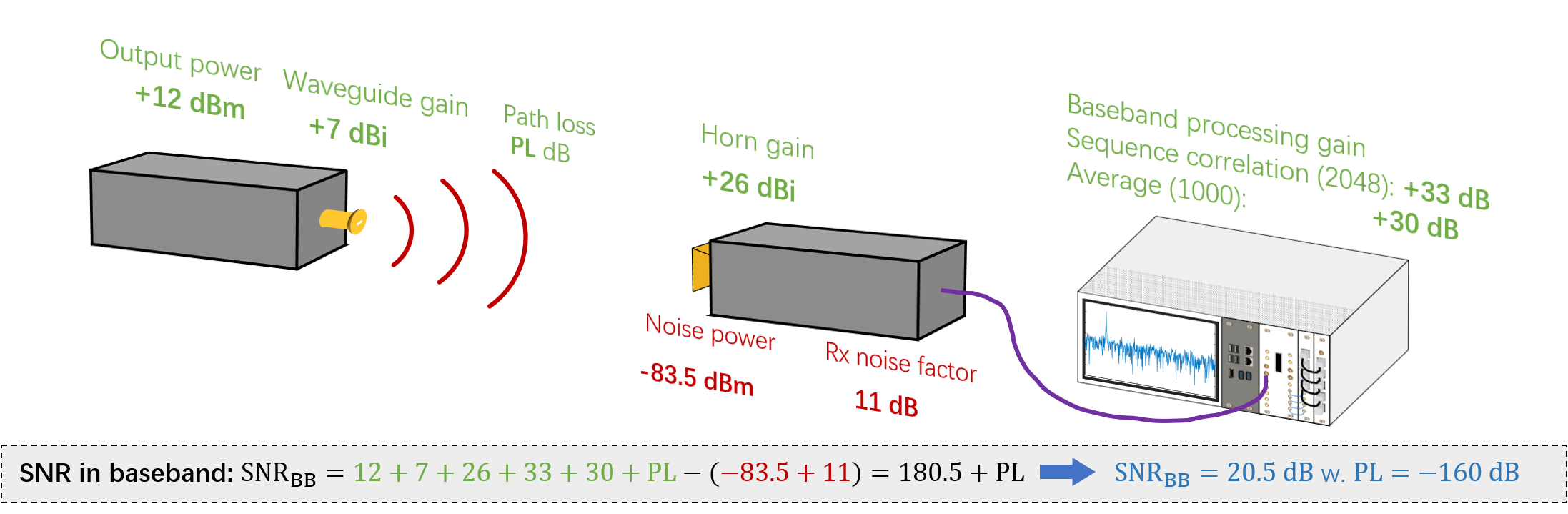}  
    \caption{The link budget analysis considering the measurement setup in this work. Note that the Rx gains are omitted here as they affect both the signal and noise and thus cause no influence on the SNR.}
    \label{fig:budget}
    \vspace{-0.5cm}
\end{figure*}
\par The measurement setup is described in detail as follows. The measured frequency band is centered at \SI{220}{GHz}, with a bandwidth of \SI{1.536}{GHz}. Consequently, the delay resolution is \SI{0.65}{ns}, meaning that any two MPCs with a propagation distance difference greater than \SI{19.5}{cm} can be distinguished. Additionally, the length of the ZC sequence is 2048, resulting in a maximum delay of \SI{1332.7}{ns} and a maximum path length of \SI{399.8}{m}.
\par Note that at the farthest Rx position, the Tx/Rx distance slightly exceeds the maximum path length of our sounder, resulting in circularly shifted CIR results. To address this, the CIRs measured at the last Rx position are manually extended to a length of 2154 samples, with the last 106 samples copied from the first 106 samples of the original CIR results, yielding an equivalent measurable path length of \SI{420}{m}. 
\par Moreover, the Tx radiates THz waves through a standard WR5 waveguide, which has a \SI{7}{dBi} antenna gain. In contrast, the Rx is equipped with a horn antenna featuring a \SI{26}{dBi} gain and an $8^\circ$ half-power beamwidth (HPBW). To capture MPCs from various directions, direction-scan sounding (DSS) is performed by mechanically rotating the Rx to scan the azimuth plane from $0^\circ$ to $360^\circ$ and the elevation plane from $-20^\circ$ to $20^\circ$, with a $10^\circ$ step.
\par For the link parameters, the Tx transmit power is \SI{12}{dBm}, and the Rx noise factor is \SI{11}{dB}. To mitigate the impact of noise, the measured CIRs are processed using a 1000-times averaging, providing a \SI{30}{dB} processing gain. Consequently, a link budget analysis can be performed, as shown in Fig.~\ref{fig:budget}. The signal strength at the Rx baseband is enhanced by the Tx antenna gain, Rx antenna gain, and the baseband processing gain from sequence correlation and averaging, while being attenuated by propagation path loss. Regarding noise effects, the Johnson–Nyquist noise power, calculated for a \SI{1.536}{GHz} bandwidth, is \SI{-83.5}{dBm}, which is then amplified by the Rx noise factor of \SI{11}{dB}. Note that the Rx gains from low-noise amplifiers (LNAs) are excluded from the link budget analysis, as they enhance both signal and noise strength without affecting the signal-to-noise ratio (SNR). Therefore, the received SNR in the baseband is computed as $180.5 + \text{PL}$ dB. Considering the free space path loss (FSPL) at \SI{400}{m} is approximately \SI{130}{dB}, a \SI{30}{dB} dynamic range is achieved while maintaining an SNR of \SI{20}{dB}, demonstrating the effectiveness of our channel sounder for long-distance measurements.
\subsection{Data Post-processing}
\begin{figure}[!tbp]
    \centering
    \subfloat[Time drift test.]{
    \centering
    \includegraphics[width=0.8\columnwidth]{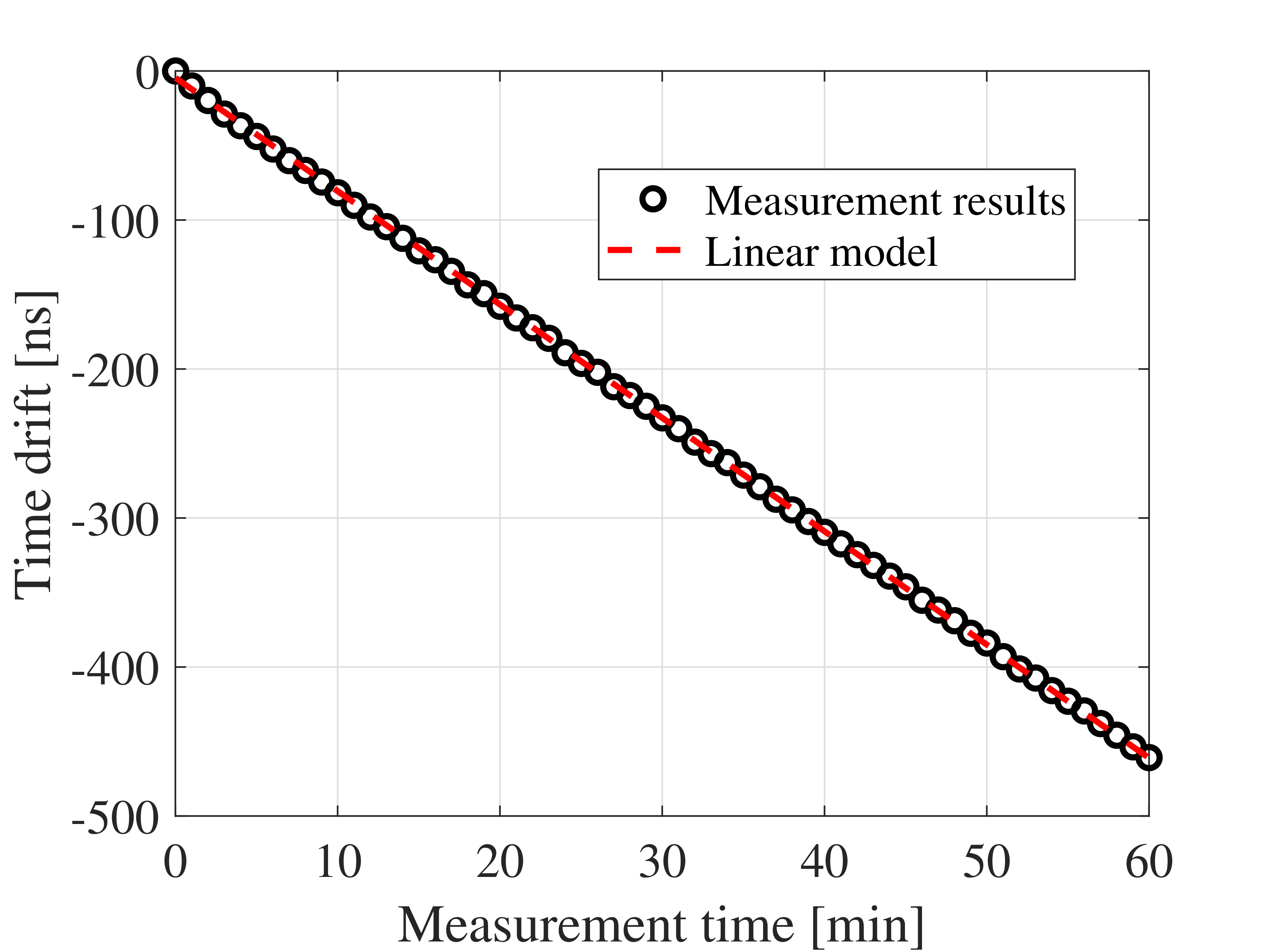}  
    }
    \quad
    \subfloat[Time drift in outdoor measurements.]{ 
    \centering
    \includegraphics[width=0.8\columnwidth]{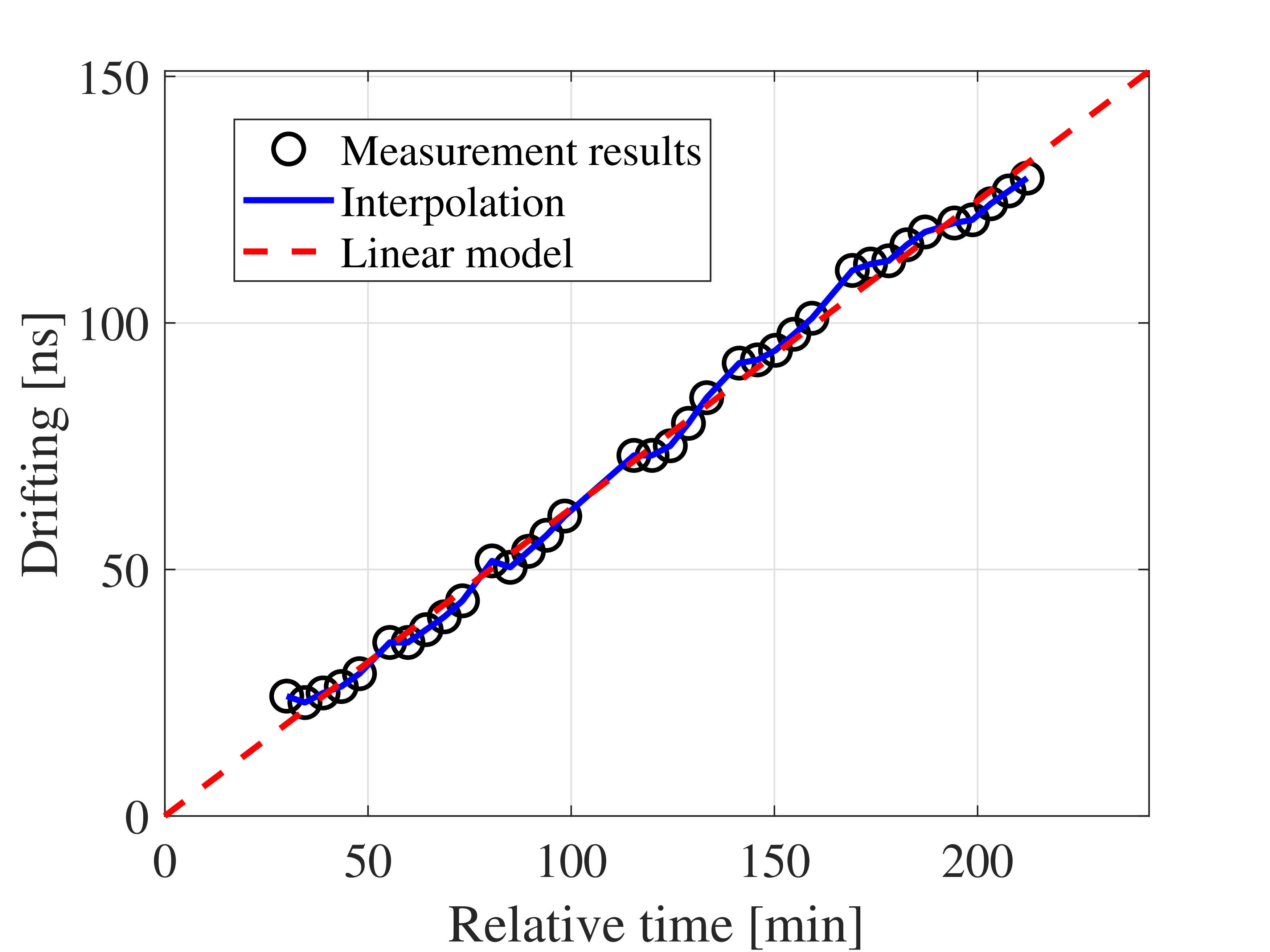}  
    }
    \caption{The time drifts between the Tx and Rx.}
    \label{fig:tdc}
    \vspace{-0.5cm}
\end{figure}
\par For accurate analysis and modeling in the THz UMa, the measured raw CIR data must undergo careful processing, which includes calibration, time-drift correction, channel estimation, and MPC clustering. First, a back-to-back calibration process is performed to eliminate the system response. Although the two Rb clocks of the Tx and Rx subsystems are synchronized before the measurements, the two 1PPS signals gradually drift apart over time. The time drift typically follows a linear relationship with respect to the measurement time, as shown in Fig.~\ref{fig:tdc}(a). The time drift at any given time can be predicted by a linear model fitted with known drift values at specific reference time instances\cite{MacCartney2017Flexible}. The simple time drift test in Fig.~\ref{fig:tdc}(a) shows a \SI{1.22}{ns} root-mean-square error (RMSE) between the measured time drift and the linear model, which is sufficient to support the long-distance channel measurements in this work. However, in outdoor environments, the rate of time drift may also vary over time, as shown in Fig.~\ref{fig:tdc}(b), potentially due to significant temperature fluctuations between day and night. Therefore, a linear interpolation method is preferred to fit the time drift curve in segments, as detailed in our previous work\cite{Li2023Correlation}.
\par Furthermore, as the spatial profiles are measured through DSS, the antenna response influences the measured results, which is mitigated by accurate channel parameter estimation using the DSS-o-SAGE algorithm~\cite{li2024sage}. Additionally, to analyze the cluster parameters in THz channels, the MPCs are further grouped into clusters using a multipath component distance (MCD)-based density-based spatial clustering of applications with noise (DBSCAN) algorithm, demonstrating strong clustering performance~\cite{chen2021channel}. To optimize processing time, only MPCs with path gains exceeding a predefined threshold are extracted during the channel parameter estimation process. The path gain threshold is set as
\begin{equation}
    \alpha_\text{th} = 10^{(\max{(\alpha_1-\text{R},\text{NF}+20)}/20)},
\end{equation}
where $\alpha_1$ represents the path gain of the strongest path. $\text{R}$ denotes the dynamic range, which is set to \SI{30}{dB} in this work. Moreover, $\text{NF}$ refers to the average noise floor of the measured CIRs, which is \SI{-180.5}{dB} for our sounder. Note that the DSS-o-SAGE algorithm may generate spurious MPCs from noise samples if the path gain threshold is too close to the noise floor. To mitigate this, a \SI{20}{dB} range is applied to account for noise amplitude fluctuations and the Rx antenna gain. Additionally, $\max{(\text{A},\text{B})}$ represents the maximum value between A and B.
\section{Propagation Analysis and Channel Characterization}
\par In this section, we analyze the propagation phenomena and channel characteristics in the THz UMa. First, the dominant propagation paths in the THz UMa are identified, and their corresponding physical causes are discussed. We also examine the primary objects that influence THz wave propagation. Additionally, we calculate and analyze key channel characteristics, including path loss, shadow fading, K-factor, delay spread, angular spread, and cluster parameters.
\label{sec:char}
\begin{figure*}[!tbp]
    \centering
    \subfloat[PDPs in LoS/OLoS.]{
    \centering
    \includegraphics[width=0.85\columnwidth]{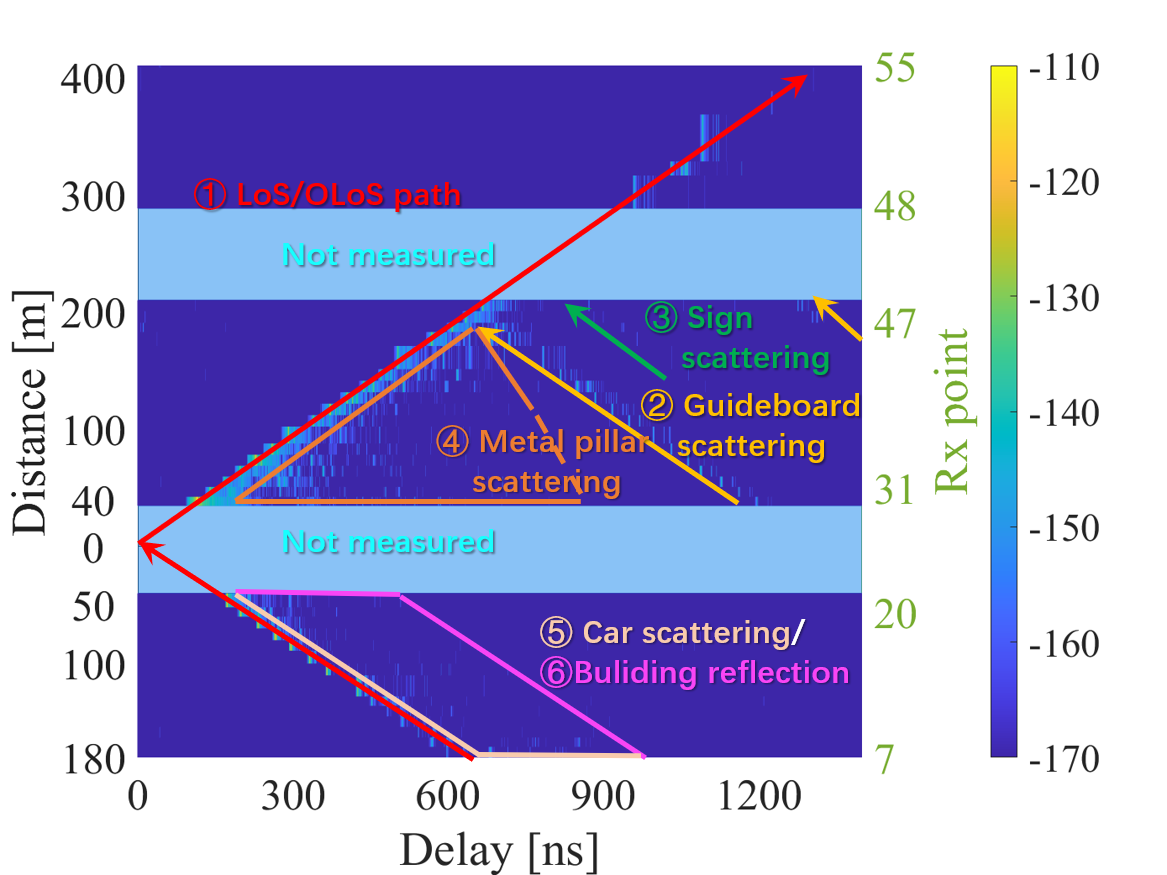}  
    }
    \subfloat[PDPs in NLoS.]{ 
    \centering
    \includegraphics[width=0.85\columnwidth]{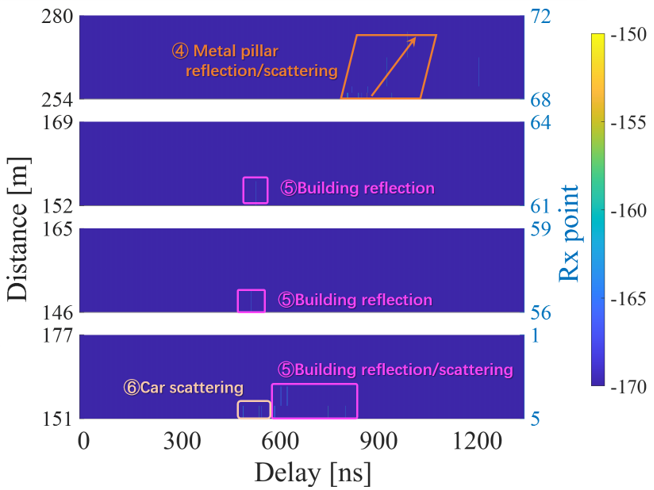}  
    }
    \quad
    \subfloat[Objects.]{ 
    \centering
    \includegraphics[width=0.85\columnwidth]{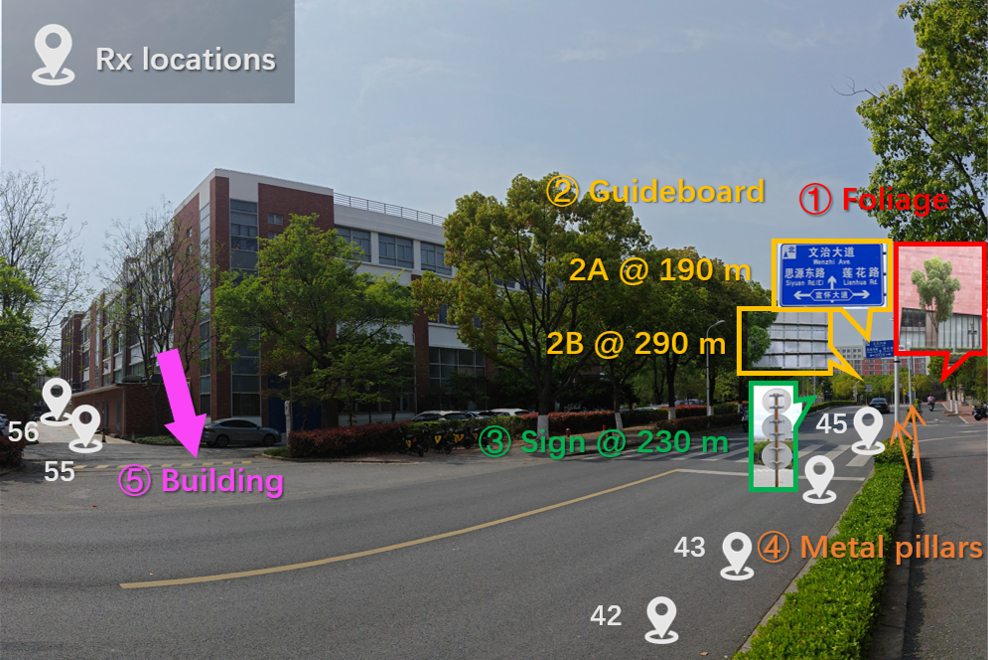}  
    }
    \subfloat[Objects.]{ 
    \centering
    \includegraphics[width=0.85\columnwidth]{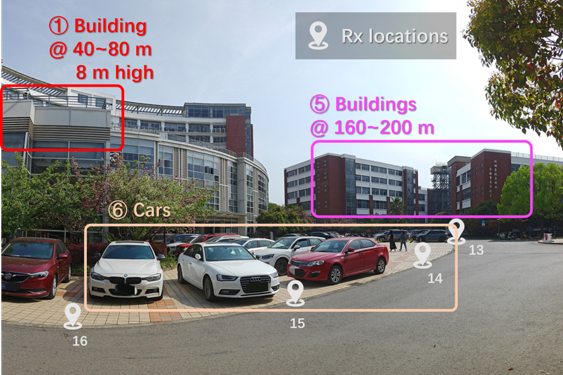}  
    }
    \caption{The propagation analysis in the THz UMa. PDPs measured in different Rx locations are shown in (a) and (b), while the objects affecting the THz propagation are shown in (c) and (d).}
    \label{fig:sp}
    \vspace{-0.5cm}
\end{figure*}

\subsection{Propagation Analysis}
\label{sec:scandpa}
\par To analyze the propagation phenomena, the PDPs at different Rx locations are combined and presented in Fig.~\ref{fig:sp}. The PDPs at each Rx location are obtained by summing the PDPs measured in all directions. Noise samples, i.e., those with power below \text{NF}, are set to a near-zero value (e.g., -200 dB) to prevent an increase in the noise floor due to the superimposition of multiple PDPs. The PDPs for LoS/OLoS and NLoS conditions are shown in Fig.~\ref{fig:sp}(a) and (b), respectively, with typical objects highlighted in Fig.~\ref{fig:sp}(c) and (d), based on images taken near Rx 16 and Rx 42, respectively.
\par Several observations can be made, as follows. First, in the LoS/OLoS case, the evolution of the LoS/OLoS path along the movement of the Rx is clearly observable. As the propagation delay of the LoS/OLoS path is positively correlated with the Tx-Rx distance, the trajectory of the LoS/OLoS path forms a straight line with a slope corresponding to the speed of light. It is worth noting that the blockage of the LoS path occurs due to different reasons for Rx 7–20 and Rx 31–55. The former are affected by an 8 m high building, as shown in Fig.~\ref{fig:sp}(d), while the latter are obstructed by foliage. Second, for Rxs along Nanyangdong Road, several trajectories in the PDPs originate from back-scattering paths from guideboards and signs. As the Tx-Rx distance increases, the propagation delay of these paths decreases, eventually disappearing as the Rx moves farther than the scatterers. The positions of the guideboards and signs are indicated in Fig.~\ref{fig:sp}(c), corresponding to the evolution of the scattering paths. Among these scattering paths, the one from guideboard 2A is significant, while others are much weaker as they stem from the backs of the guideboards and signs. Fourth, metal pillars and cars near the Rx cause rich scattering along the Rx route. However, the signal strength is weak, and these scattering events appear randomly with insignificant spatial consistency. Fifth, considering the 20 Rx locations in the NLoS case, it can be observed from Fig.~\ref{fig:sp}(b) that those Rx locations deep in NLoS areas, i.e., those far from the turning point from LoS to NLoS, barely receive any effective signals. For the remaining Rx locations, the THz channels are sparse, with only a few paths originating from car scattering, metal pillar scattering, and building reflection/scattering. This observation suggests that coverage extension techniques may be necessary to ensure effective network performance in the THz UMa.
\subsection{Path Loss and Shadow Fading}
\begin{figure}
    \centering
    \includegraphics[width=0.8\columnwidth]{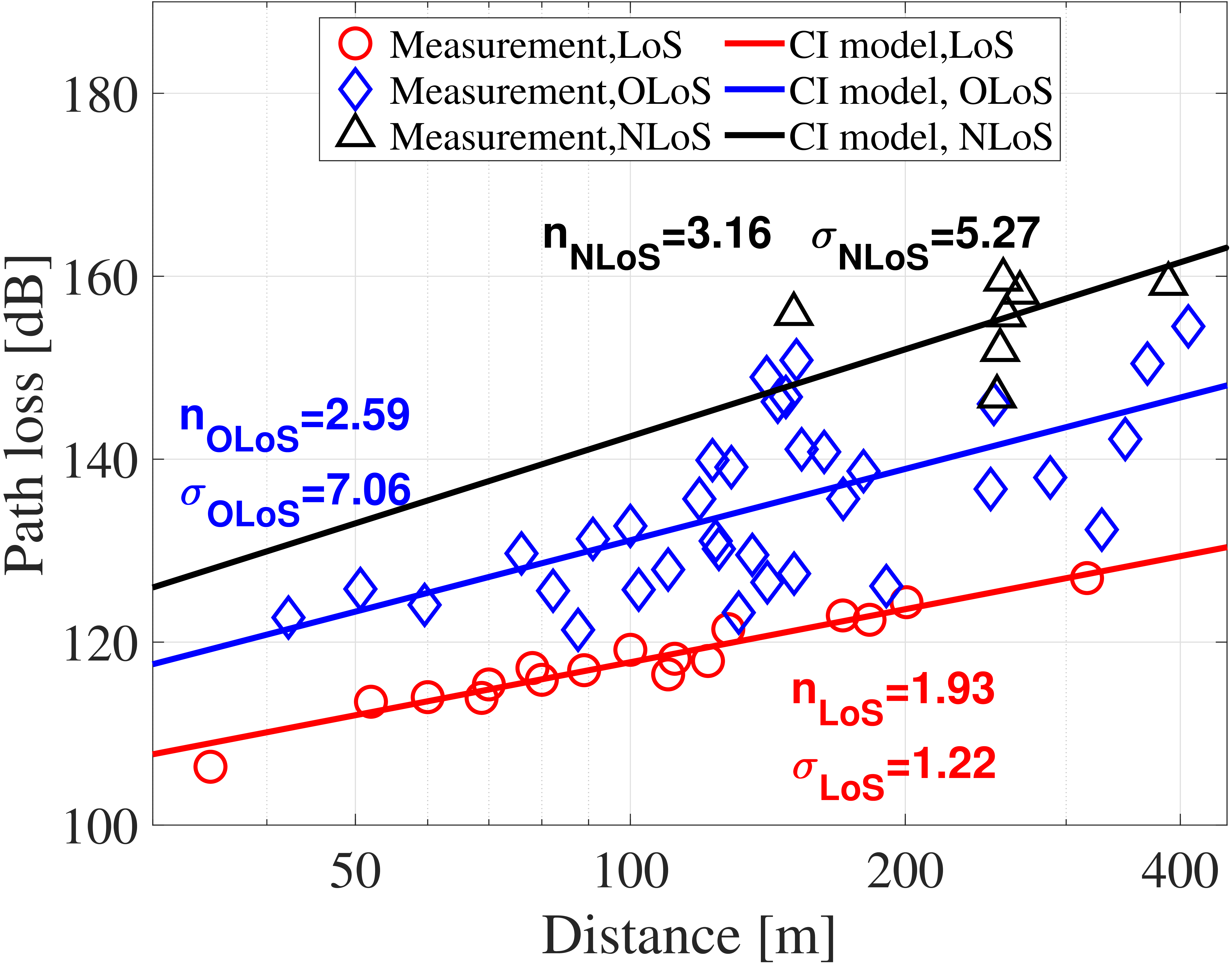}
    \caption{Path loss and shadow fading results in the THz UMa.}
    \label{fig:plandsf}
\end{figure}
\par Known as large-scale fadings, path loss and shadow fading are key parameters for link budget analysis. The path loss is related to the overall received power from all propagation paths, which is fitted using the close-in (CI) free space reference distance model, expressed as
\begin{equation}
    \text{PL}^\text{CI}~[\text{dB}]=10n\log_{10}\frac{d}{d_0}+\text{FSPL}(f_0,d_0)+\chi,
\end{equation}
where $n$ denotes the path loss exponent (PLE). Moreover, $d_0$ is the reference distance that is set as \SI{1}{m} in this work, while $f_0$ is the center frequency of \SI{220}{GHz}. $\text{FSPL}(f_0,d_0)$ stands for the FSPL at the corresponding frequency and distance, which is calculated using the Friss' equation, as
\begin{equation}
    \text{FSPL}(f,d)~[\text{dB}]=20\log_{10}\frac{c}{4\pi fd},
\end{equation}
where $c$ denotes the speed of light. Moreover, the shadow fading is termed as a zero-mean Gaussian distributed random variable $\chi$. 
\par Several observations are made, as follows. First, the measured PLEs are 1.93, 2.59, and 3.16 for the LoS, OLoS, and NLoS cases, respectively. The PLE in the LoS case is close to 2, which is the typical PLE for free space with only the LoS path, indicating that the NLoS paths in the THz UMa are relatively weak. Second, weak shadow fading effects are observed in the LoS case, with a standard deviation of \SI{1.22}{dB}. In contrast, the shadowing effect is more pronounced in the OLoS and NLoS cases, with standard deviations of \SI{7.06}{dB} and \SI{5.27}{dB}, respectively.
\subsection{K-factor}
\begin{figure}
    \centering
    \includegraphics[width=0.9\columnwidth]{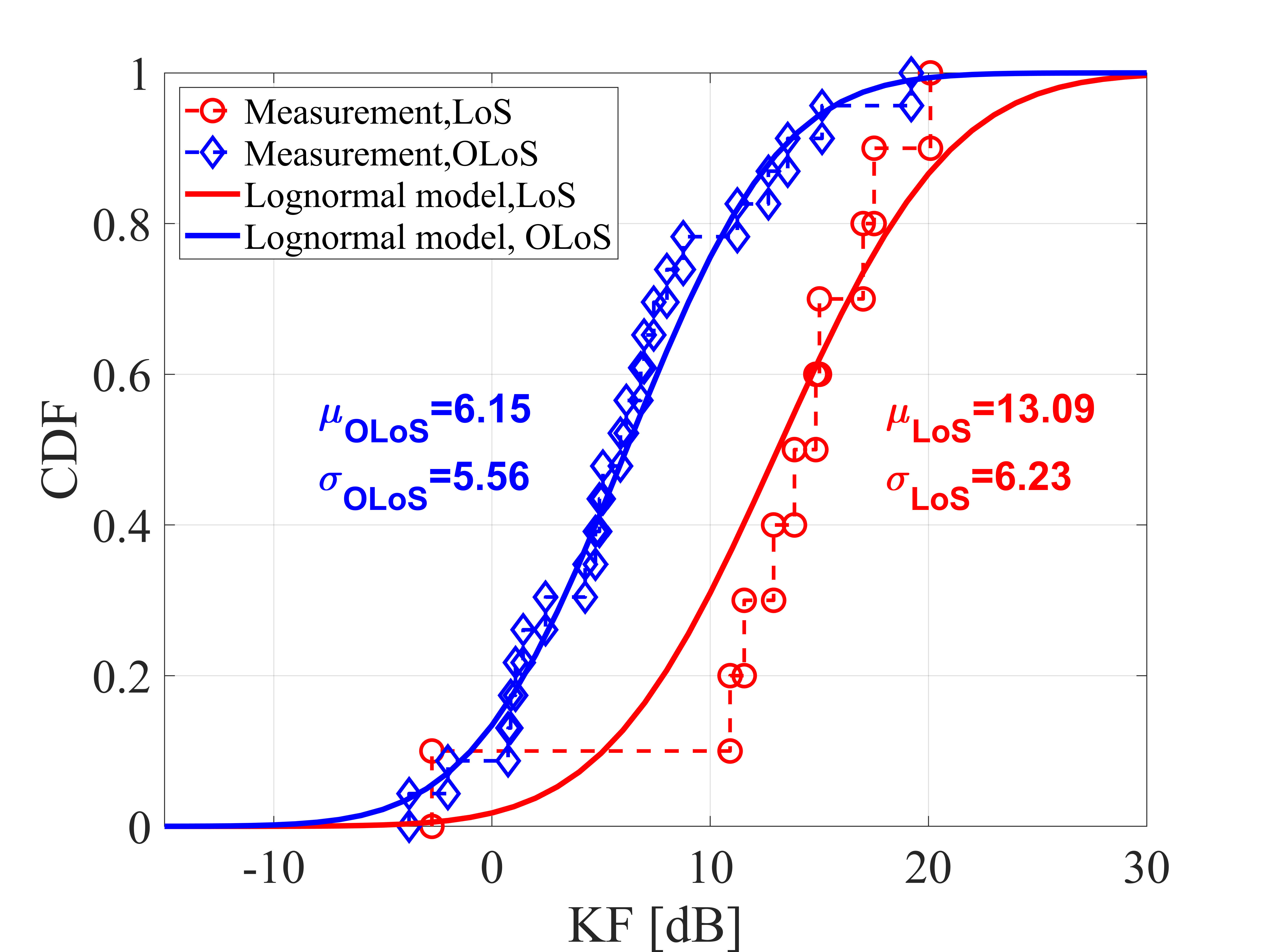}
    \caption{KF results and fitting models in the THz UMa.}
    \label{fig:kf}
\end{figure}
\par The K-factor quantifies the dominance of the strongest cluster, with the results fitted to a log-normal distribution, as shown in Fig.~\ref{fig:kf}. Note that in most NLoS locations, only one cluster, or even no cluster, is observed, leading to insufficient K-factor values for model fitting in the NLoS cases, which are thus omitted here. From the results in Fig.~\ref{fig:kf}, the mean K-factor values are \SI{13.09}{dB} and \SI{6.15}{dB} for the LoS and OLoS cases, respectively. The higher K-factor values in the LoS case indicate the strong dominance of the LoS cluster. In the OLoS case, the THz channels remain primarily dominated by the OLoS cluster, but the K-factor values are smaller than those in the LoS case, as the OLoS cluster power is attenuated by obstruction losses from foliage and buildings.

\subsection{Delay Spread}
\begin{figure}
    \centering
    \includegraphics[width=0.9\columnwidth]{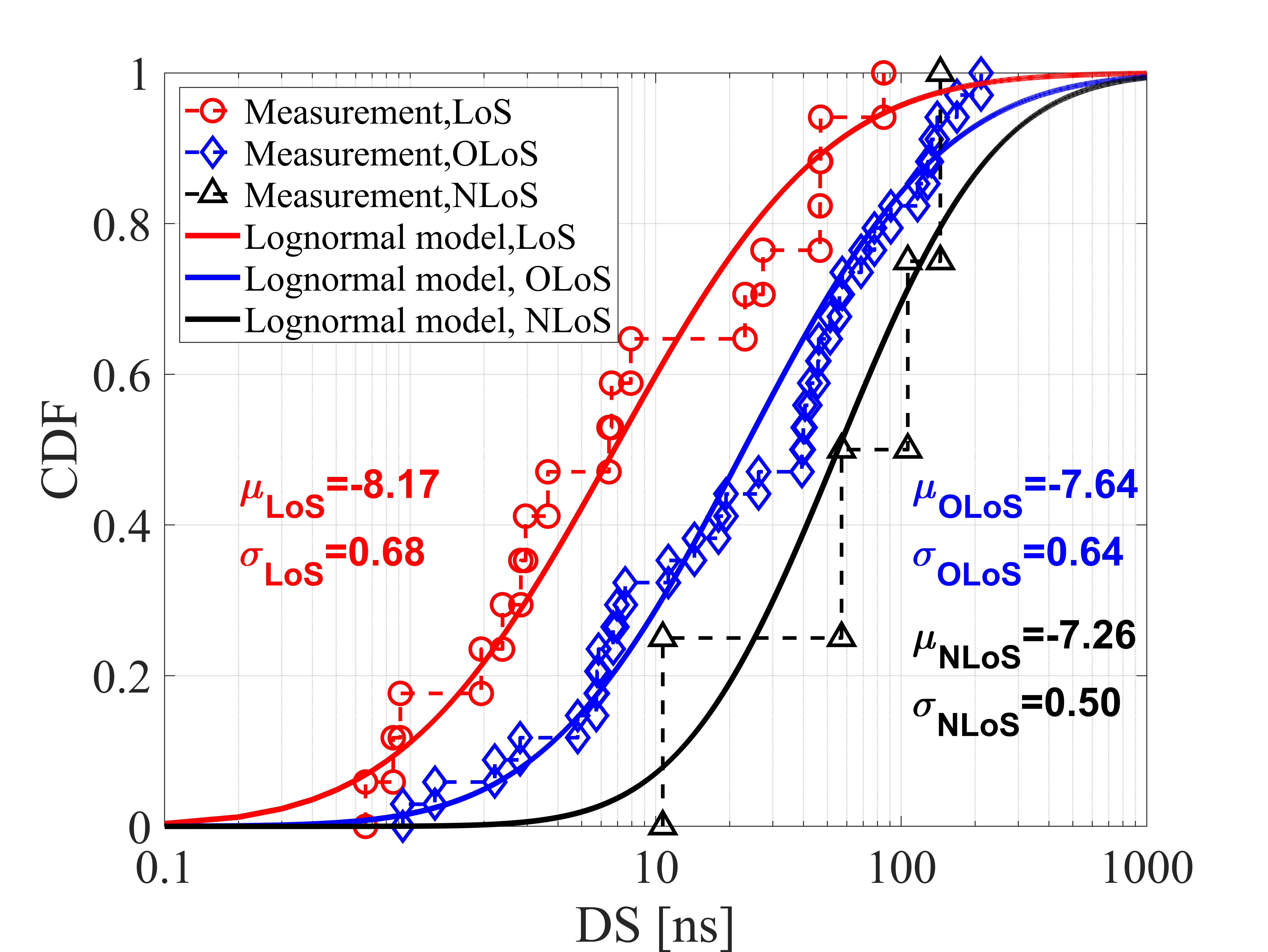}
    \caption{DS results and fitting models in the THz UMa.}
    \label{fig:ds}
\end{figure}
\par The power of MPCs disperses in the temporal domain, which may lead to inter-symbol interference (ISI) and must be carefully considered in communication system designs. This dispersion can be characterized by the root-mean-square (RMS) delay spread. The measured delay spread values are fitted to log-normal distributions, as shown in Fig.~\ref{fig:ds}. The average delay spread values are \SI{6.76}{ns}, \SI{22.9}{ns}, and \SI{54.95}{ns} for the LoS, OLoS, and NLoS cases, respectively. These delay spreads are larger than those reported in our previous work on THz picocell scenarios, where the DS values are typically several nanoseconds in the LoS case and up to \SI{20}{ns} in the NLoS case\cite{wang2023thz, li2024pico}. This discrepancy is attributed to the higher base station height in this work, allowing for the reception of MPCs with larger delays, such as scattering paths from guideboards and signs, which contribute to the larger delay spread values. Additionally, the delay spread values increase as the power of the LoS path decreases, transitioning from the LoS case to the OLoS and NLoS cases. This trend is expected, as the influence of other NLoS paths becomes more prominent when the LoS path power diminishes.
\subsection{Angular Spreads}
\begin{figure}[!tbp]
    \centering
    \subfloat[ASA.]{ 
    \centering
    \includegraphics[width=0.85\columnwidth]{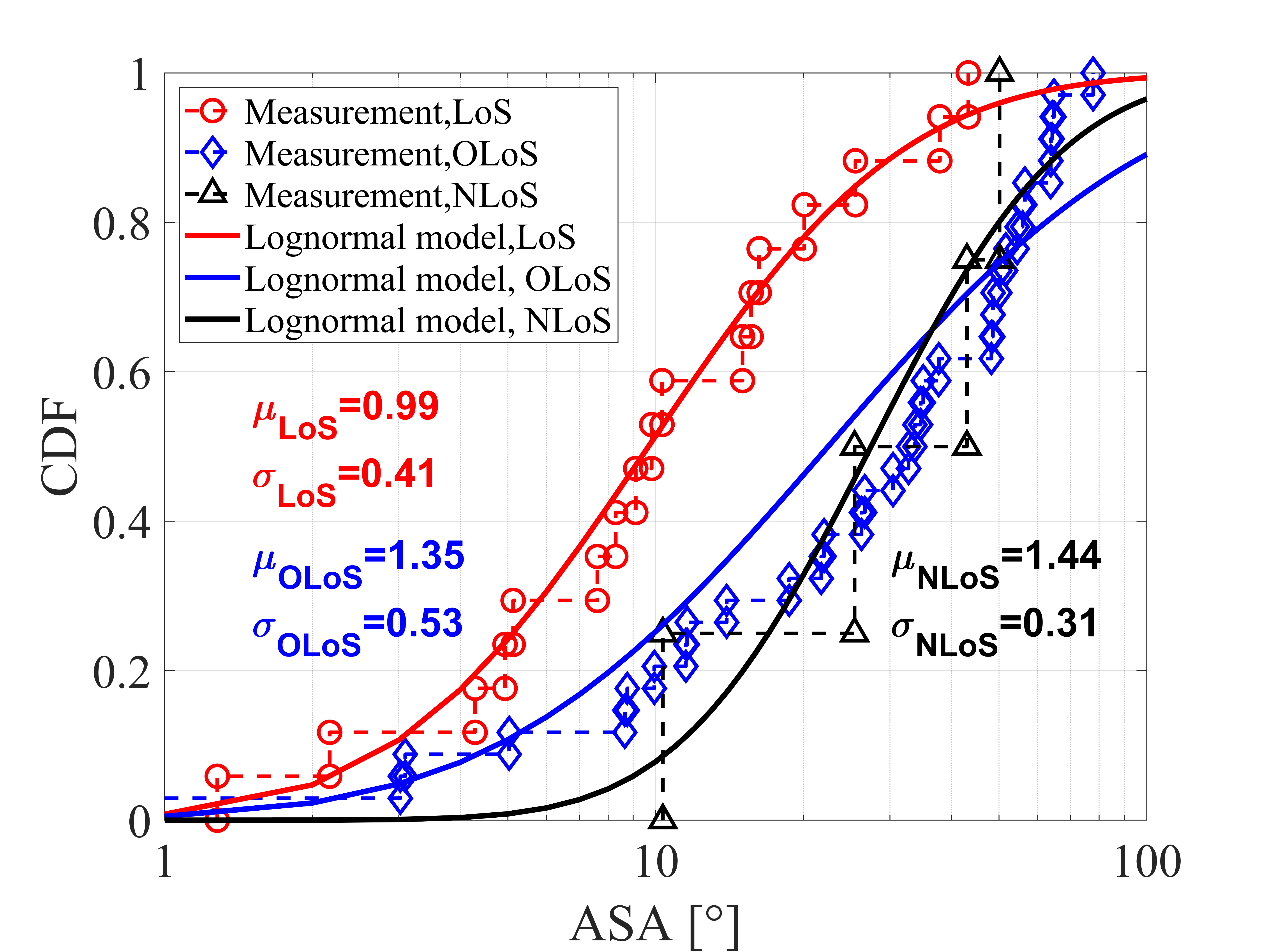}  
    }
    \quad
    \subfloat[ESA.]{ 
    \centering
    \includegraphics[width=0.85\columnwidth]{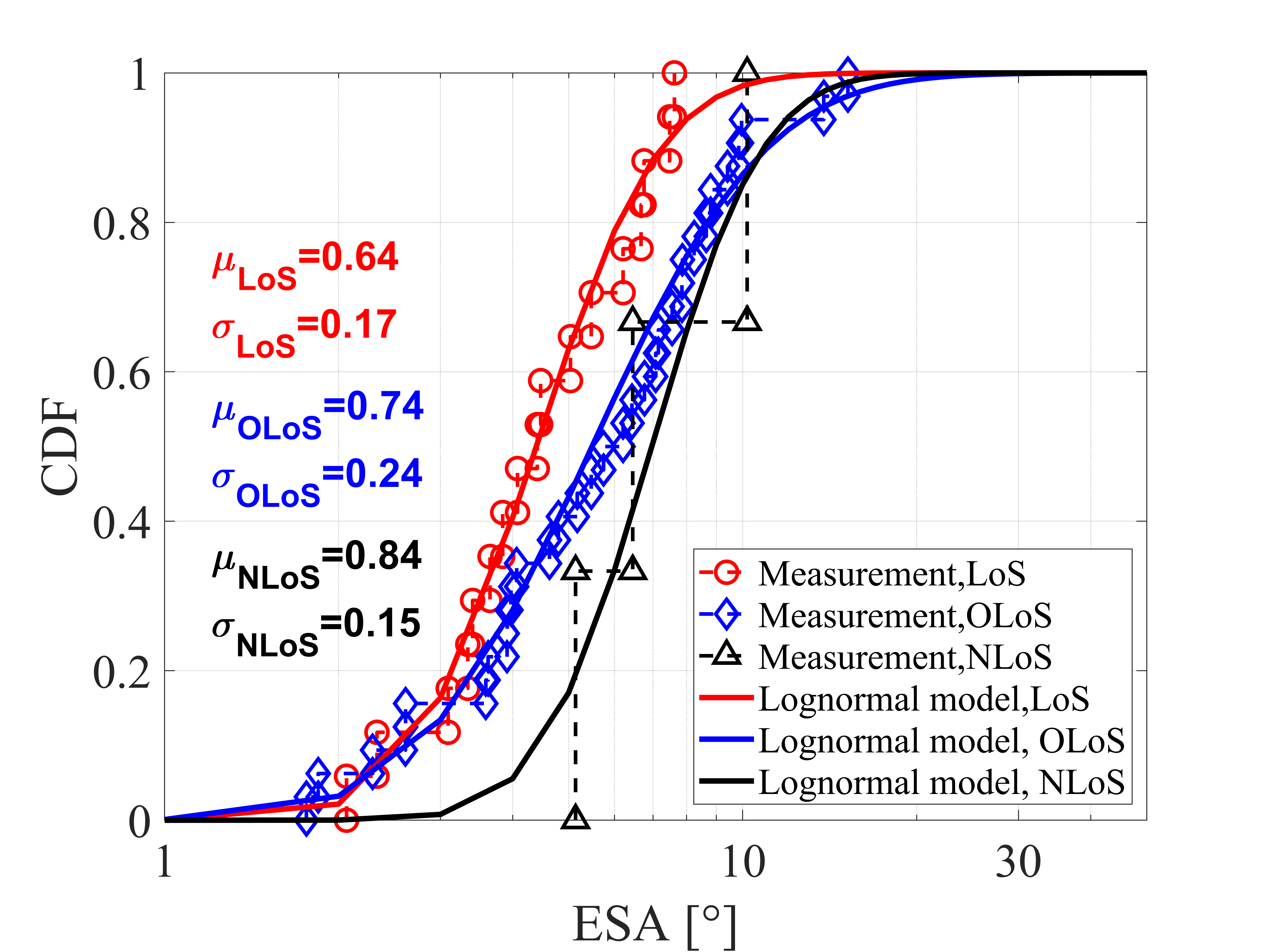}  
    }
    \caption{The angular spread values in the THz UMa}
    \label{fig:as}
    \vspace{-0.5cm}
\end{figure}
\par Similar to the delay spread, the power of the multipath components (MPCs) disperses in the spatial domain, which is characterized by the angular spreads. The azimuth spread of arrival (ASA) and elevation spread of arrival (ESA) results are calculated and fitted to a log-normal distribution, as shown in Fig.~\ref{fig:as}. The average ASA values are $9.77^\circ$, $22.38^\circ$, and $27.54^\circ$ for the LoS, OLoS, and NLoS cases, respectively, while the mean ESA values are $4.37^\circ$, $5.50^\circ$, and $6.92^\circ$ for the LoS, OLoS, and NLoS cases, respectively. Similar to the observation from the delay spread results, stronger multipath effects are observed as the LoS path power decreases.
\subsection{Cluster Parameters}
\par The cluster parameters, including the number of clusters, cluster delay spread (CDS), cluster azimuth spread of arrival (CASA), and cluster elevation spread of arrival (CESA), are calculated and discussed as follows. First, the average numbers of clusters are 2.12, 2.94, and 1 for the LoS, OLoS, and NLoS cases, respectively, indicating a strong sparsity in the THz UMa scenarios. Interestingly, the number of clusters is the largest in the OLoS case. This can be attributed to two factors. On one hand, due to the large propagation loss in the NLoS case, the number of clusters is very limited. On the other hand, as we set a constant dynamic range with respect to the power of the strongest path during data processing, the number of NLoS clusters included within this range is smaller in the LoS case than in the OLoS case, owing to the stronger LoS path power. Second, the values for cluster delay and angular spreads are as follows: \SI{2.83}{ns}, \SI{3.95}{ns}, and \SI{3.23}{ns} for delay spread; $1.72^\circ$, $4.27^\circ$, and $6.23^\circ$ for ASA; and $3.58^\circ$, $4.01^\circ$, and $2.47^\circ$ for ESA, for the LoS, OLoS, and NLoS cases, respectively. Unlike the delay and angular spread results, there is no obvious trend with respect to the LoS states, as the cluster spreads are highly dependent on the shape of the clusters, which is determined by the types and sizes of scatterers, rather than the LoS conditions.
\section{Digital Twin Enabled Hybrid Channel Modeling}
\label{sec:dtecm}
\begin{figure*}[!tbp]
    \centering
    \includegraphics[width=2.0\columnwidth]{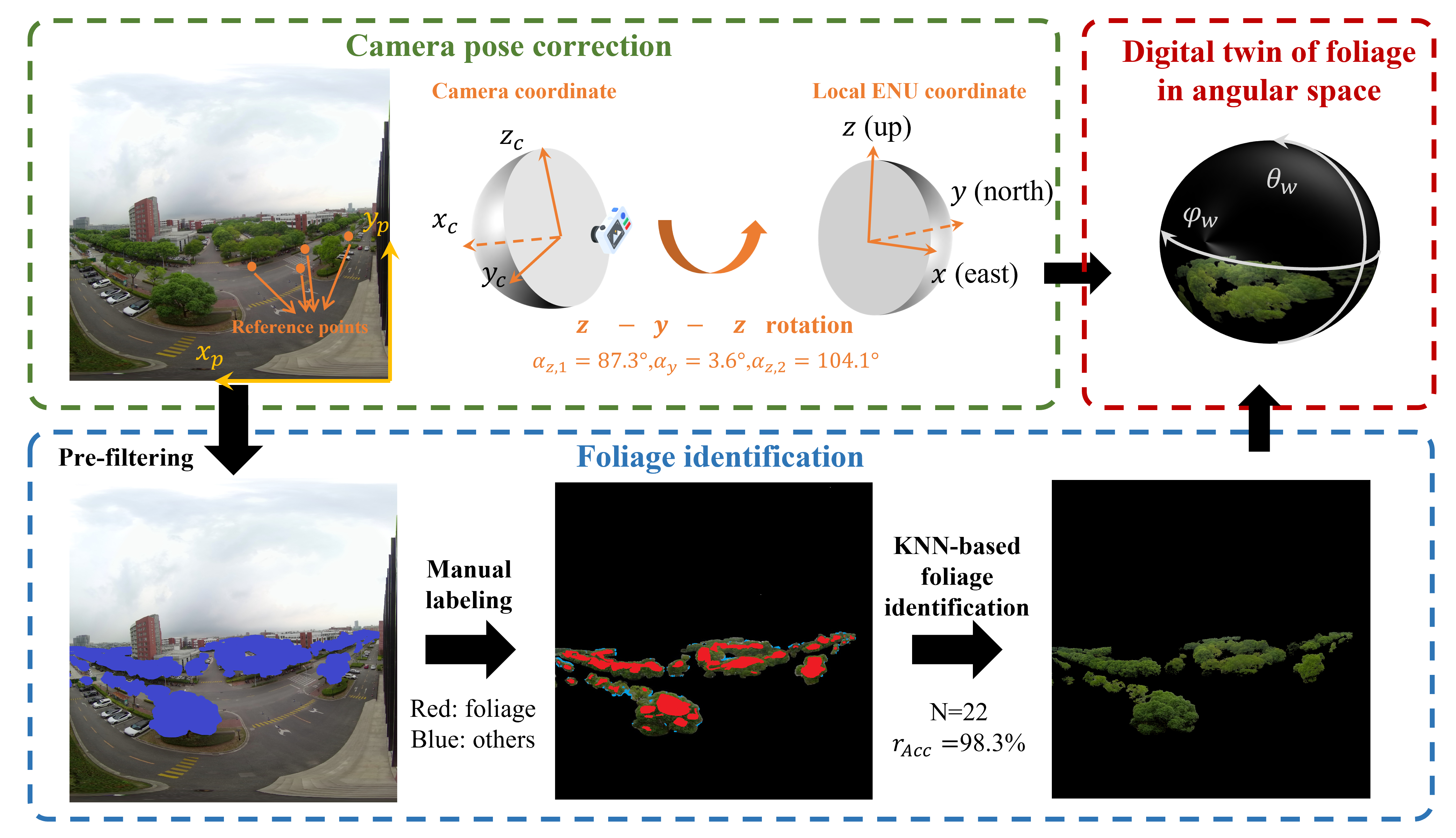}  
    \caption{The creation steps of the digital twin of foliage.}
    \label{fig:foliagedt}
    \vspace{-0.5cm}
\end{figure*}
\par In this section, the DTECM is introduced. First, to accurately model the additional loss due to foliage blockage, a digital twin of the foliage, namely the accurate angular positions of foliage, is generated from a panoramic image using a KNN-based foliage identification method and camera pose correction. The foliage loss is then modeled by a segmented linear model based on the FCR of the Tx beam in the angular domain. Furthermore, a hybrid channel model is presented, which combines ray-tracing, CV techniques, and statistical methods. Finally, the performance of the hybrid model is validated and compared with the existing 3GPP standard channel model. Link performance, in terms of spectral efficiency (SE) and coverage ratio, is evaluated using the DTECM in the THz UMa.
\subsection{Digital Twin of Foliage}
\par As discussed previously, foliage significantly influences propagation loss in the THz UMa. Therefore, to accurately capture this effect, we create a digital twin of the foliage in the angular space to aid in effective channel modeling at the THz band. The creation of the digital twin involves camera pose correction and foliage identification, as illustrated in Fig.~\ref{fig:foliagedt}, which is explained in detail as follows.
\subsubsection{Camera pose correction}
To obtain the digital twin of the foliage, a panoramic image is captured with the camera placed at the same position as the Tx. The panoramic camera maps the objects in the camera's spherical coordinate system to the rectangular image using equirectangular projection (ERP), where the projection relationship is
\begin{align}
    \varphi_c&=x_c\Delta\varphi +\varphi_{c,0},\\
    \theta_c&=y_c\Delta\theta+\theta_{c,0},
\end{align}
where $\Delta\varphi$ and $\Delta\theta$ are the angular resolution in the azimuth and elevation planes, respectively. Moreover, $\varphi_{c,0}$ and $\theta_{c,0}$ are the corresponding angles of the right-bottom pixel in the camera coordinate system. 
\par When capturing the panoramic image, the camera is placed arbitrarily, causing the camera's coordinate system to deviate from the local east-north-up (ENU) system, as illustrated in Fig.~\ref{fig:foliagedt}. Specifically, the x-axis in the camera coordinate system, i.e., $\varphi_c = \theta_c = 0$, aligns with the camera's steering direction, while the y- and z-axes correspond to the left and upward directions on the camera lens surface. In contrast, for channel analysis and modeling, the widely adopted coordinate system is the local ENU system, which is centered at the Tx.
\par Therefore, a coordinate rotation is required to map the objects captured in the panoramic image to the local ENU coordinate system. This is achieved through a $z-y-z$ rotation process. Specifically, we perform sequential rotations of the camera coordinate system by an angle $\alpha$ around the $z$ axis, an angle $\beta$ around the $y$ axis, and an angle $\gamma$ around the $z$ axis to align it with the local ENU coordinate system. To determine the rotation angles, we select several reference points with known coordinates, as shown in Fig.~\ref{fig:foliagedt}. The optimal rotation angles are found by minimizing the distance between the rotated coordinates in the camera system and the true local ENU coordinates, as given by
\begin{equation}
    \begin{split}
        [\alpha_{z,1},\alpha_y,\alpha_{z,2}]=
        \text{arg}\min_{\alpha,\beta\,\gamma}\sum_{i=1}^I\left|[\varphi_w^\text{i},\theta_w^\text{i}]-f^{z-y-z}_{\alpha,\beta,\gamma}(\varphi_c^i,\theta_c^i)\right|^2,
    \end{split}
\end{equation}
where $I$ denotes the number of reference points. $\varphi_w^i$ and $\varphi_c^i$ are the azimuth angles of the $i^\text{th}$ reference point in the local ENU coordinate and the camera coordinate, respectively, while $\theta_w^i$ and $\theta_c^i$ are the elevation angles of the $i^\text{th}$ reference point in the local ENU coordinate and the camera coordinate, respectively. Moreover, the rotation functions can be expressed as concatenations of the rotation functions around the z-axis and y-axis, as
\begin{equation}
    f^{z-y-z}_{\alpha,\beta,\gamma}(\varphi,\theta)=f_\gamma^z(f_\beta^y(f_\alpha^z(\varphi,\theta))),
\end{equation}
where the rotation functions around the z- and y-axis are expressed as
\begin{align}
    f_\alpha^z(\varphi,\theta)&=[\varphi+\alpha,\theta]\\
    f_\alpha^y(\varphi,\theta)&=[\arctan(\frac{\cos\theta\sin\varphi}{\cos\alpha\cos\theta\cos\varphi+\sin\alpha\sin\theta}),\\
    &~~~\arcsin(-\sin\alpha\cos\theta\cos\varphi+\cos\alpha\sin\theta)].
\end{align}
\par Through a simple exhaustive search process, the rotation angles are found to be $\alpha_{z,1}=87.3^\circ$, $\alpha_y=3.6^\circ$, $\alpha_{z,2}=104.1^\circ$.
\subsubsection{Foliage identification}
To model the foliage loss, the foliage location information must be extracted from the panoramic image, for which a KNN-based foliage identification method is employed. Given the large size of the panoramic image (3840 × 3840 pixels), directly processing the entire image would be computationally expensive and time-consuming. To reduce computational complexity, we first perform a manual pre-filtering of the image by roughly highlighting the foliage areas with a reference color. The foliage regions are then approximately identified using a color difference metric, defined as
\begin{equation}
    \bm{S}_{\text{Foliage}}=\left\{(x_c,y_c);\Delta C(\text{RGB}[x_c,y_c],\text{RGB}_{\text{ref}})<\Delta C_{\text{th}}\right\},
\end{equation}
where $x_c$ and $y_c$ denote the pixel-wise locations in the picture. Moreover, $\text{RGB}[x,y]$ denotes the RGB vector for the pixel in the $x$ raw and $y$ column, while $\text{RGB}_{\text{ref}}$ stands for the RGB vector of the selected color, which is (63,71,204) in this work. $\Delta C_{\text{th}}$ is the threshold parameter, which is set as $50$ here. Additionally, the color difference metric is expressed as
\begin{equation}
\begin{split}
        \Delta C(C_1,C_2)=~~~~~~~~~~~~~~~~~~~~~~~~~~~~~~~~~~~~~&\\
        \sqrt{(2+\frac{\overline{r}}{256})\Delta R^2+4\Delta G^2+(2+\frac{255-\overline{r}}{256})\Delta B^2},
        \label{eq:colordiff}
\end{split}
\end{equation}
where the intermediate variables are calculated as
\begin{align}
    \overline{r}&=\frac{C_{1,R}+C_{2,R}}{2},\\
    \Delta R&=C_{1,R}-C_{2,R},\\
    \Delta G&=C_{1,G}-C_{2.G},\\
    \Delta B&=C_{1,B}-C_{2.B},
\end{align}
where $C_{1/2,R/G/B}$ denotes the red/green/blue color values of the two colors.
\par
\begin{figure}
    \centering
    \includegraphics[width=0.8\columnwidth]{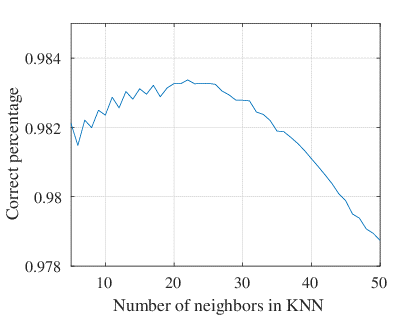}
    \caption{The accuracy of the KNN method for foliage identification versus number of neighbors.}
    \label{fig:knn}
    \vspace{-0.5 cm}
\end{figure}

 Furthermore, to accurately extract the foliage from the rough foliage areas, the KNN method is employed to differentiate between foliage and other objects based on their color characteristics. To train the KNN algorithm, pixels are manually labeled, with red pixels representing foliage and blue pixels indicating non-foliage objects. RGB value sets for foliage and non-foliage objects are then obtained by collecting the RGB values from the corresponding red and blue areas. The KNN method is subsequently applied using the color difference in~\eqref{eq:colordiff} as the distance metric to determine whether a given pixel in the foliage area corresponds to foliage or another object. Since the number of neighbors, $N$, influences the performance of the KNN algorithm, repeated experiments are conducted to evaluate the algorithm's accuracy for varying $N$ values. Specifically, 60$\%$ of the labeled data are used as training sets, while the remaining 40$\%$ serve as test sets, with the results presented in Fig.~\ref{fig:knn}.
As shown in Fig.~\ref{fig:knn}, the accuracy of the KNN method for foliage identification is notably high (over 98$\%$), confirming its effectiveness. Furthermore, as $N$ increases, the classification accuracy initially rises and then declines. The highest accuracy of $98.3\%$ is achieved for $N=22$, which is used throughout the paper. With the KNN approach, foliage can be accurately extracted from the panoramic picture, as demonstrated in Fig.~\ref{fig:foliagedt}.
\par By combining the foliage identification results with the coordinate rotation process for camera pose correction, a digital twin of the foliage is generated in the angular space, as shown in the top-right figure of Fig.~\ref{fig:foliagedt}. This digital twin enables precise modeling of foliage loss, which is described in detail in the following section.
\subsection{Foliage Loss Modeling Based on Digital Twin}
\begin{figure}
    \centering
    \subfloat[Effects of the threshold parameter.] {  
    \includegraphics[width=0.85\columnwidth]{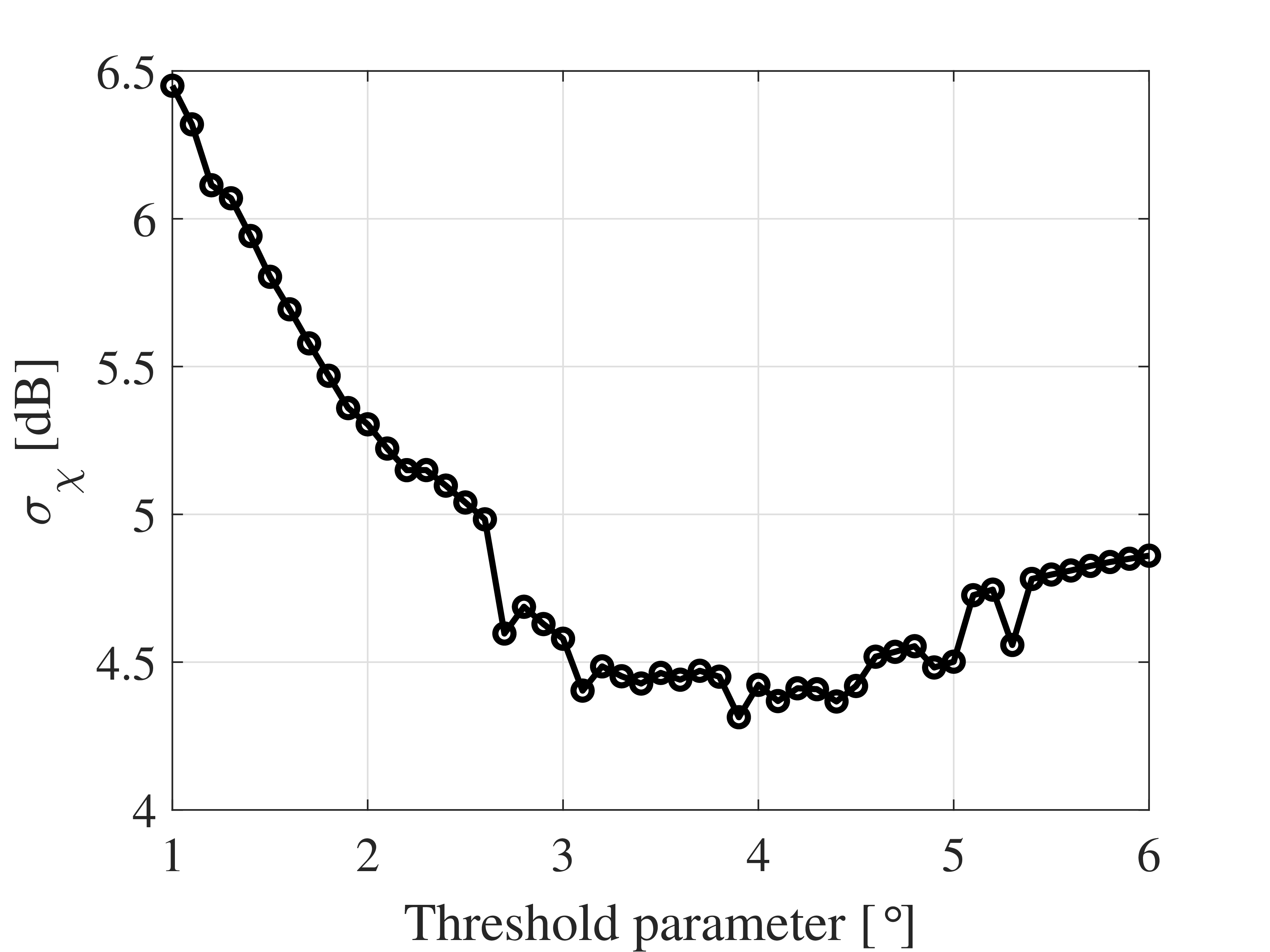}  
    }
    \quad
    \subfloat[Segmented linear model.] { 
    \includegraphics[width=0.85\columnwidth]{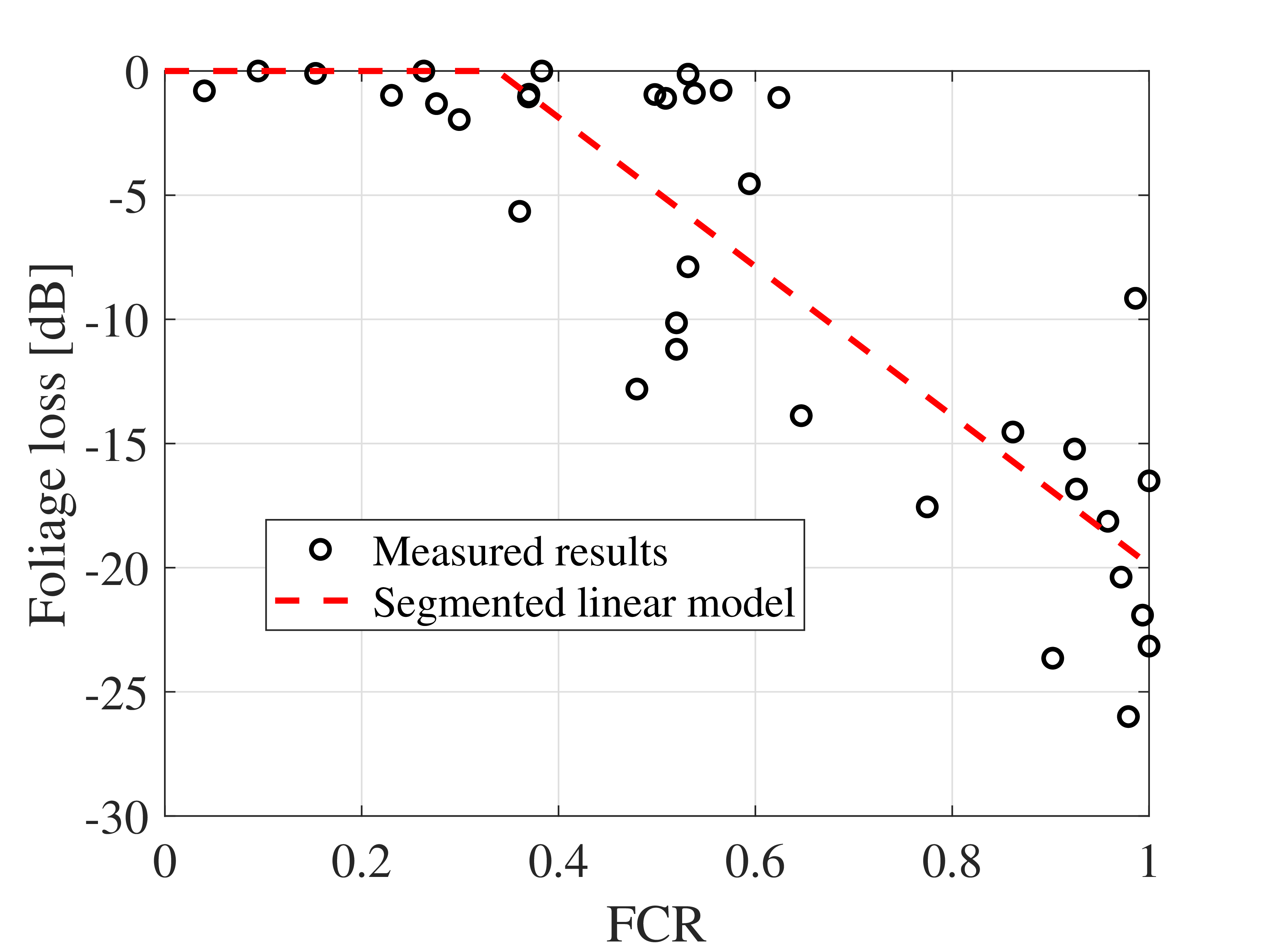}  
    }
    \caption{The segmented linear model for the foliage loss in the THz UMa.}
    \label{fig:linearFL}
    \vspace{-0.5cm}
\end{figure}
\par To model the foliage loss based on the digital twin, the concept of FCR is introduced, which is defined as
\begin{equation}
    r_{\text{fol}}(\varphi_w,\theta_w)=\frac{N_{\bm{S}(\varphi_w,\theta_w)}^\text{Fol}}{N_{\bm{S}(\varphi_w,\theta_w)}},
    \label{eq:fcr}
\end{equation}
where $N_{\bm{S}(\varphi_w,\theta_w)}^\text{Fol}$ denotes the number of pixels in the area ${\bm{S}(\varphi_w,\theta_w)}$ that belongs to foliage, while $N_{\bm{S}(\varphi_w,\theta_w)}$ is the total number of pixels in ${\bm{S}(\varphi_w,\theta_w)}$. Specifically, the area ${\bm{S}(\varphi_w,\theta_w)}$ is related to the beam footprint of the Tx, expressed as
\begin{equation}
    {\bm{S}(\varphi_w,\theta_w)}=\left\{\varphi,\theta;\sqrt{(\varphi-\varphi_w)^2+(\theta-\theta_w)^2}<\phi_{\text{th}}\right\},
\end{equation}
where $\phi_{\text{th}}$ is the threshold parameter.
\par Intuitively, as the FCR increases, a larger portion of the Tx beam becomes obstructed by foliage, leading to an increase in foliage loss. To capture this relationship, a segmented linear model is employed to model the correlation between foliage loss and FCR, namely:
\begin{equation}
    L_\text{Fol}~[\text{dB}]=
    \begin{cases}
        0,&~\text{if}~r_{\text{fol}}<r_{\text{th}}\\
        a(r_{\text{fol}}-r_{\text{th}}),&~\text{if}~r_{\text{fol}}\geq r_{\text{th}}
    \end{cases}~~+\chi_{\text{Fol}},
    \label{eq:dtfoliageloss}
\end{equation}
where $a$ and $r_{\text{th}}$ are the slope and segment point of the model, respectively. Moreover, a Gaussian distributed random variable $\chi_{\text{Fol}}$ is introduced, since the foliage loss depends not only on the FCR but also other factors, such as leave thickness, types of trees, etc.
\par Based on the measurement results, the foliage loss can be modeled, as shown in Fig~.\ref{fig:linearFL}. Specifically, the threshold parameter $\phi_\text{th}$ significantly influences the model's performance, and it is determined as the value that minimizes the standard deviation of the Gaussian variable $\chi_{\text{Fol}}$. The optimal value is found to be $3.9^\circ$, as indicated in Fig.~\ref{fig:linearFL}(a). Furthermore, the segmented linear model is fitted to the measurement results, with good fitting performance observed in Fig.~\ref{fig:foliagedt}(b). The fitting parameters are $a = -30$, $r_\text{th} = 0.337$, $\mu_\chi = -0.15\text{dB}$, and $\sigma_\chi = 4.31$ dB.
\subsection{Hybrid Channel Modeling in the THz UMa}
\begin{figure*}[!tbp]
    \centering
    \includegraphics[width=2.0\columnwidth]{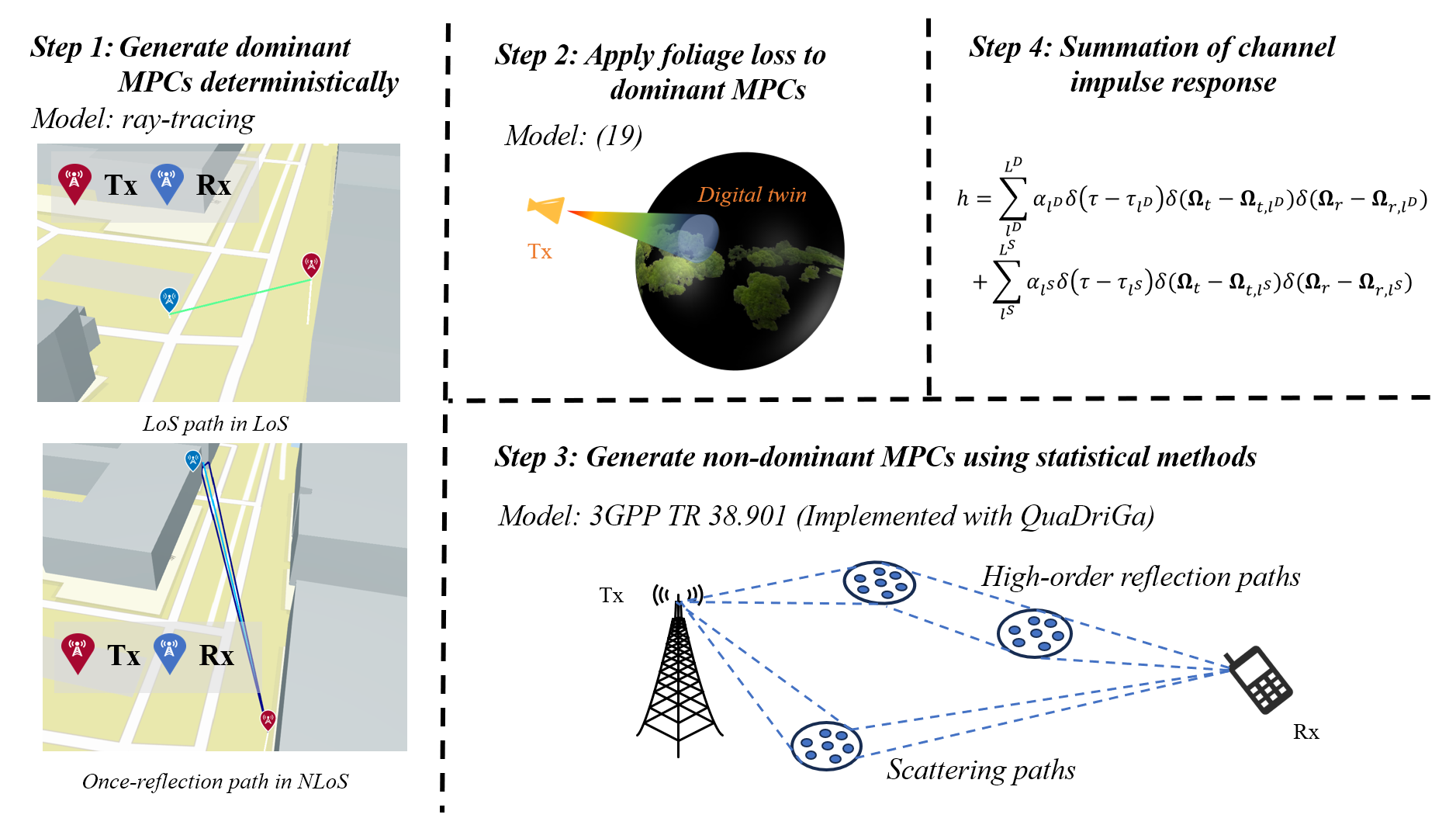}  
    \caption{The channel generation steps for the DTECM in the THz UMa.}
    \label{fig:hybridmodel}
    \vspace{-0.5cm}
\end{figure*}
\par To accurately characterize the THz wireless channels in the UMa scenario, the DTECM is proposed, combining the ray-tracing, CV technique, and statistical methods. The procedure for generating channel realizations based on the proposed DTECM is illustrated in Fig.~\ref{fig:hybridmodel} and is explained in detail as follows.
\begin{table}[tbp]
    \centering
    \caption{Fitting parameters for statistical channel generation using QuaDriGa.}
    \includegraphics[width = 0.8\columnwidth]{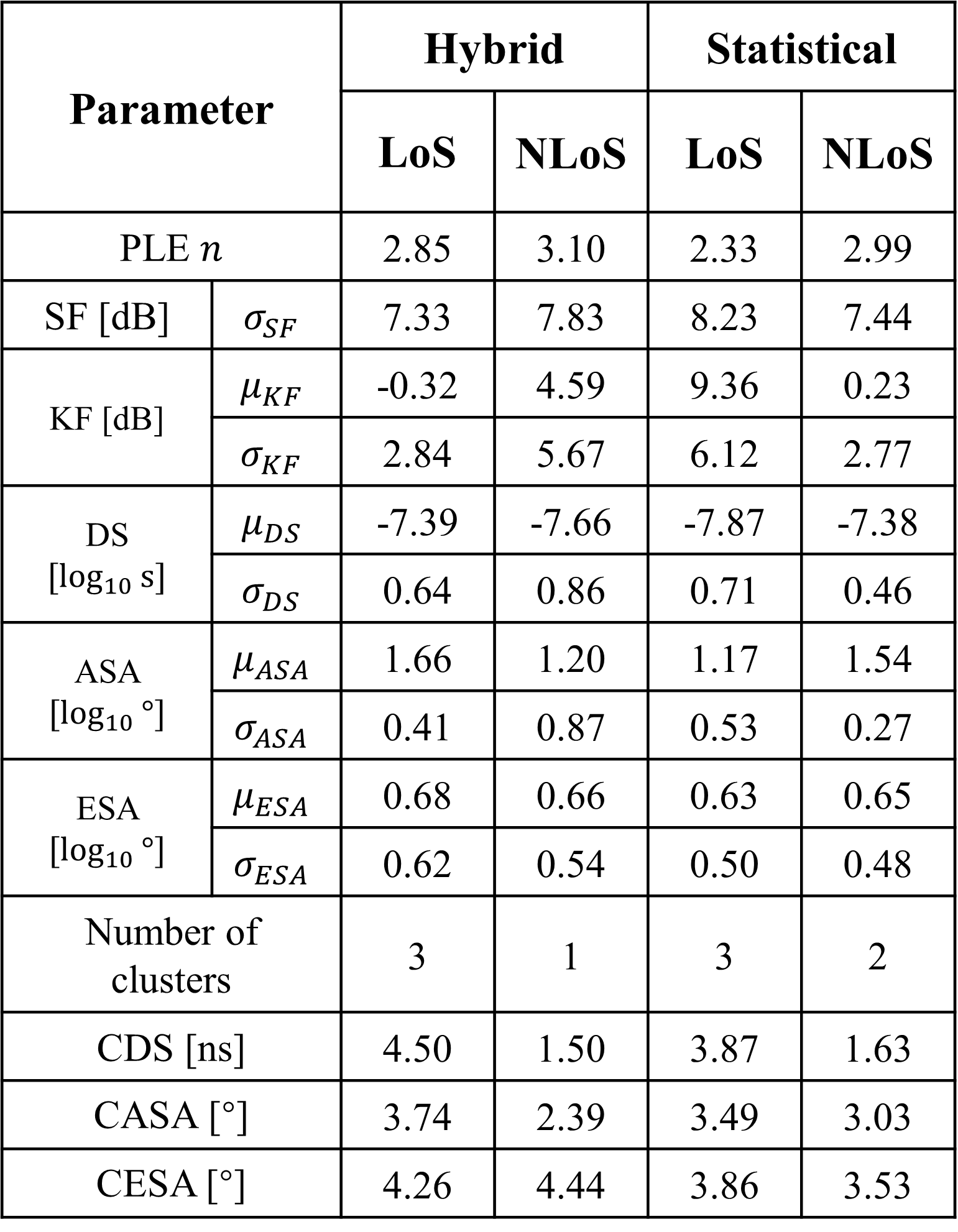}
    \label{tab:fitp}
    \vspace{-0.5cm}
\end{table}
\subsubsection{Step 1: Dominant MPC generation with ray-tracing}
\par To achieve high channel modeling accuracy, the deterministic channel modeling method, namely the ray-tracing technique, is employed. For conducting ray-tracing simulations, environment data from the measured environment are obtained from the OpenStreetMap website, which provides a free, open-source map of the world contributed by volunteers globally\footnote{www.openstreetmap.org}. It is important to note that the environment used in the ray-tracing simulation includes only the buildings in the THz UMa scenario, while the foliage is not considered at this stage. Foliage loss is incorporated in the subsequent steps. To reduce computational complexity, only the dominant paths are generated using the ray-tracing technique, including the LoS path in the LoS case and the once-reflection path in the NLoS case. It is also possible that in some deep NLoS locations, no once-reflection path is observed; such locations are considered outage locations, and the subsequent modeling steps are not applied.
\subsubsection{Step 2: Foliage loss generation for dominant paths}
\par It has been demonstrated that foliage in the THz UMa scenario can significantly impact signal propagation. Therefore, the path gains of the dominant paths generated by ray-tracing must be adjusted accordingly. This adjustment is achieved by applying the digital twin-based foliage loss model introduced in~\eqref{eq:dtfoliageloss}. For instance, for a given LoS path, once its angle of departure (AoD) is known from the ray-tracing simulation, the FCR can be calculated using~\eqref{eq:fcr}. The foliage loss is then determined using~\eqref{eq:dtfoliageloss} and applied to the path gain. A similar procedure is followed for the once-reflection path in the NLoS case, which, however, is only applicable to the path segment from the Tx to the scatterer. The foliage loss from the scatterer to the Rx is not considered here and remains a topic for future work.
\subsubsection{Step 3: Non-dominant MPCs generation with statistical methods}
\par Since only the once-reflection paths are generated by the ray-tracing simulation to reduce computational complexity, other MPCs, such as scattering paths, are generated using the statistical method based on the 3GPP channel standardization file, specifically 3GPP TR 38.901~\cite{3gpp.38.901}. The implementation of the 3GPP standardization is carried out using the quasi-deterministic radio channel generator (QuaDriGa), developed by the Fraunhofer Heinrich-Hertz-Institut (HHI)~\cite{QuaDriGa}. To effectively apply QuaDriGa to our specific scenario, it is parameterized according to the channel characterization distribution parameters provided in Table~\ref{tab:fitp}, which are derived by eliminating the effects of the dominant paths from our measurement results.
\subsubsection{Step 4: Channel impulse response calculation}
With the aforementioned three steps, the channel parameters of the MPCs, including path gain, delay, AoD, angle of arrival (AoA), and others, are determined. Based on these parameters, the channel impulse response can be constructed by combining the dominant MPCs generated deterministically and the non-dominant MPCs generated statistically. In this manner, the DTECM achieves a favorable balance between high modeling accuracy and low computational complexity, the effectiveness of which is validated in the subsequent section.
\subsection{Performance Validation and Comparison}
\begin{figure}
    \centering
    \subfloat[LoS.] {   
    \includegraphics[width=0.8\columnwidth]{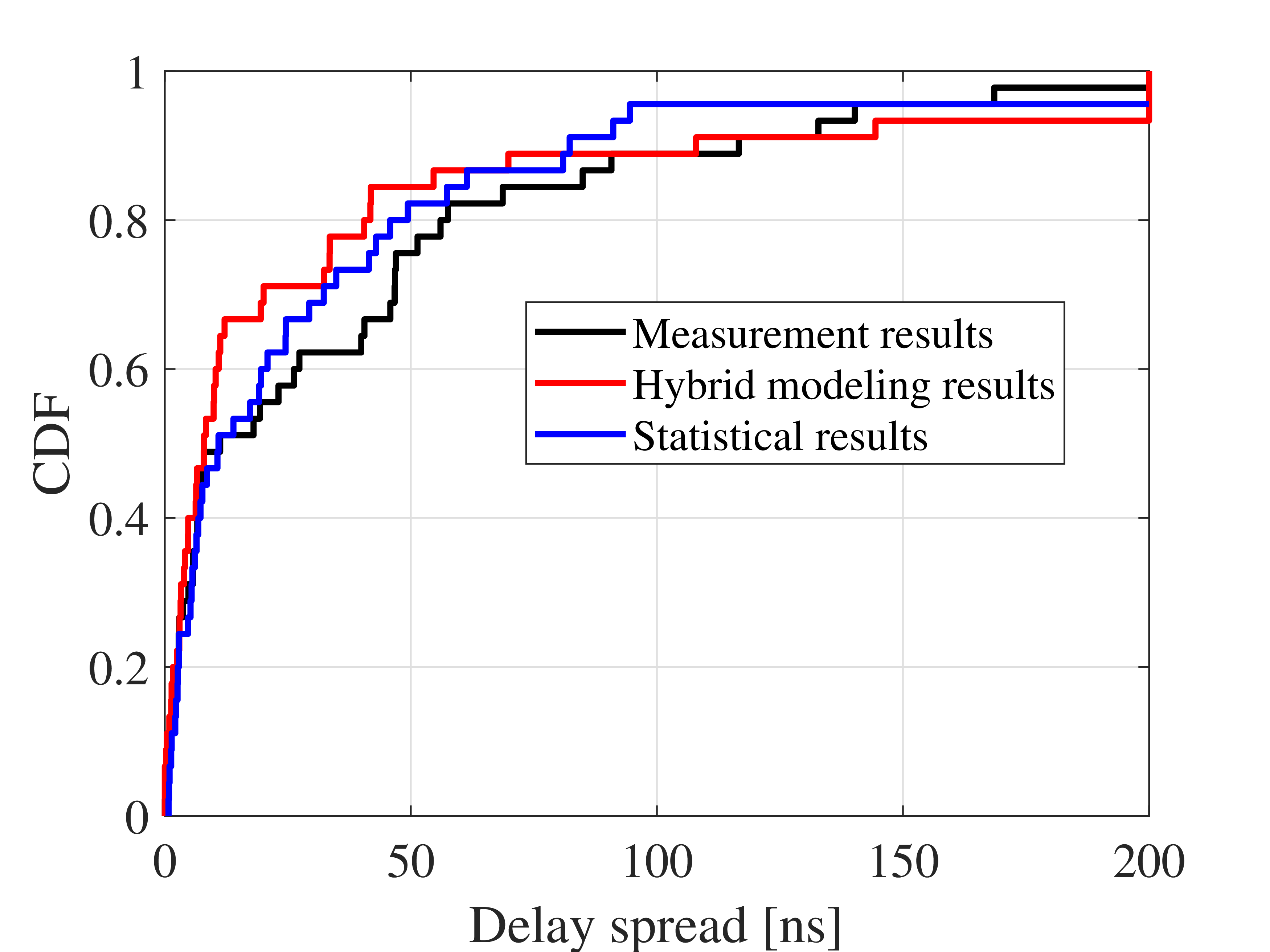}  
    }
    \quad
    \subfloat[NLoS.] {
    \includegraphics[width=0.8\columnwidth]{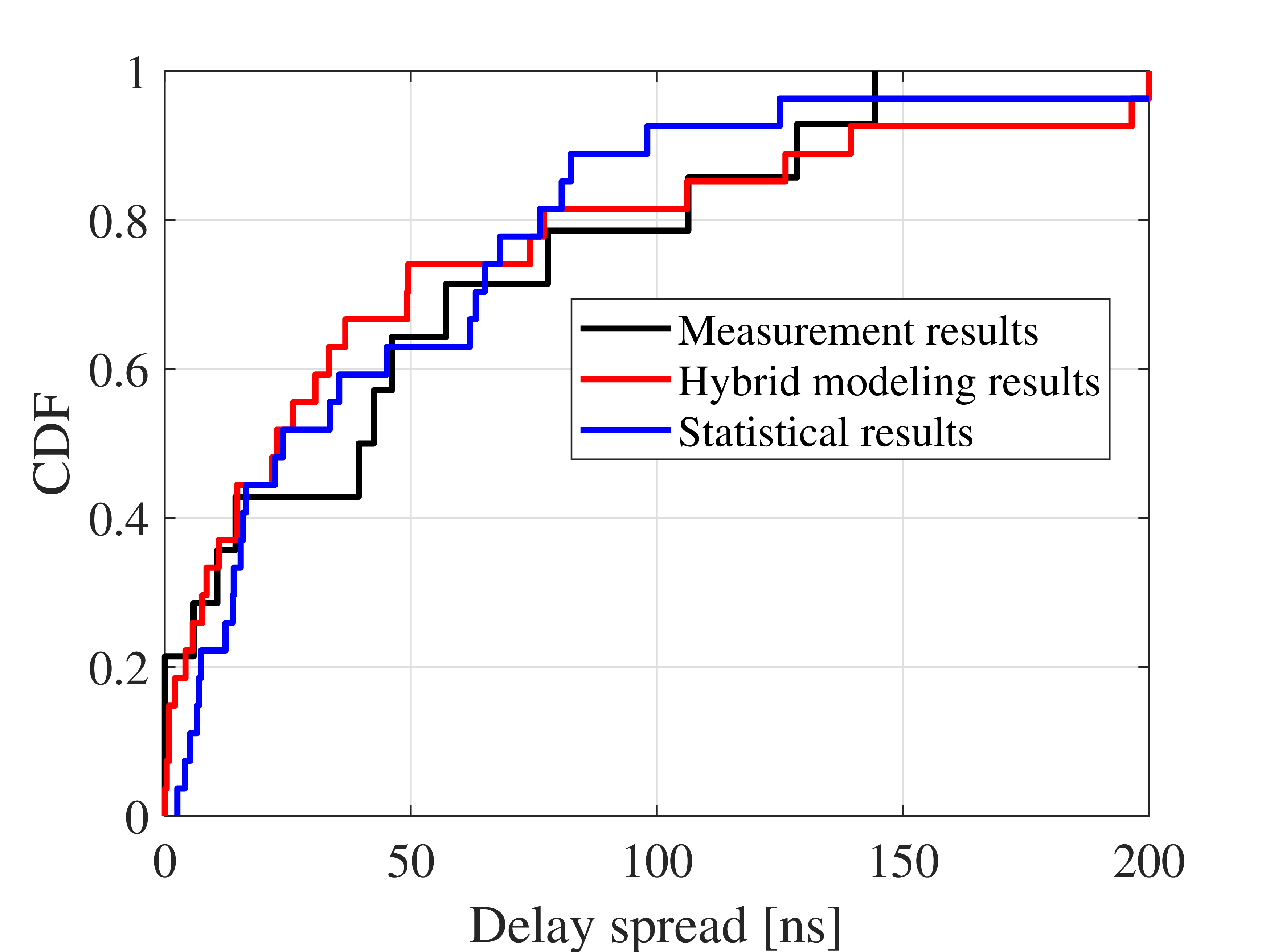}  
    }
    \caption{The delay spread results with different channel models.}
    \label{fig:dscomp}
    \vspace{-0.5cm}
\end{figure}
\begin{figure}
    \centering
    \subfloat[LoS.] { 
    \includegraphics[width=0.8\columnwidth]{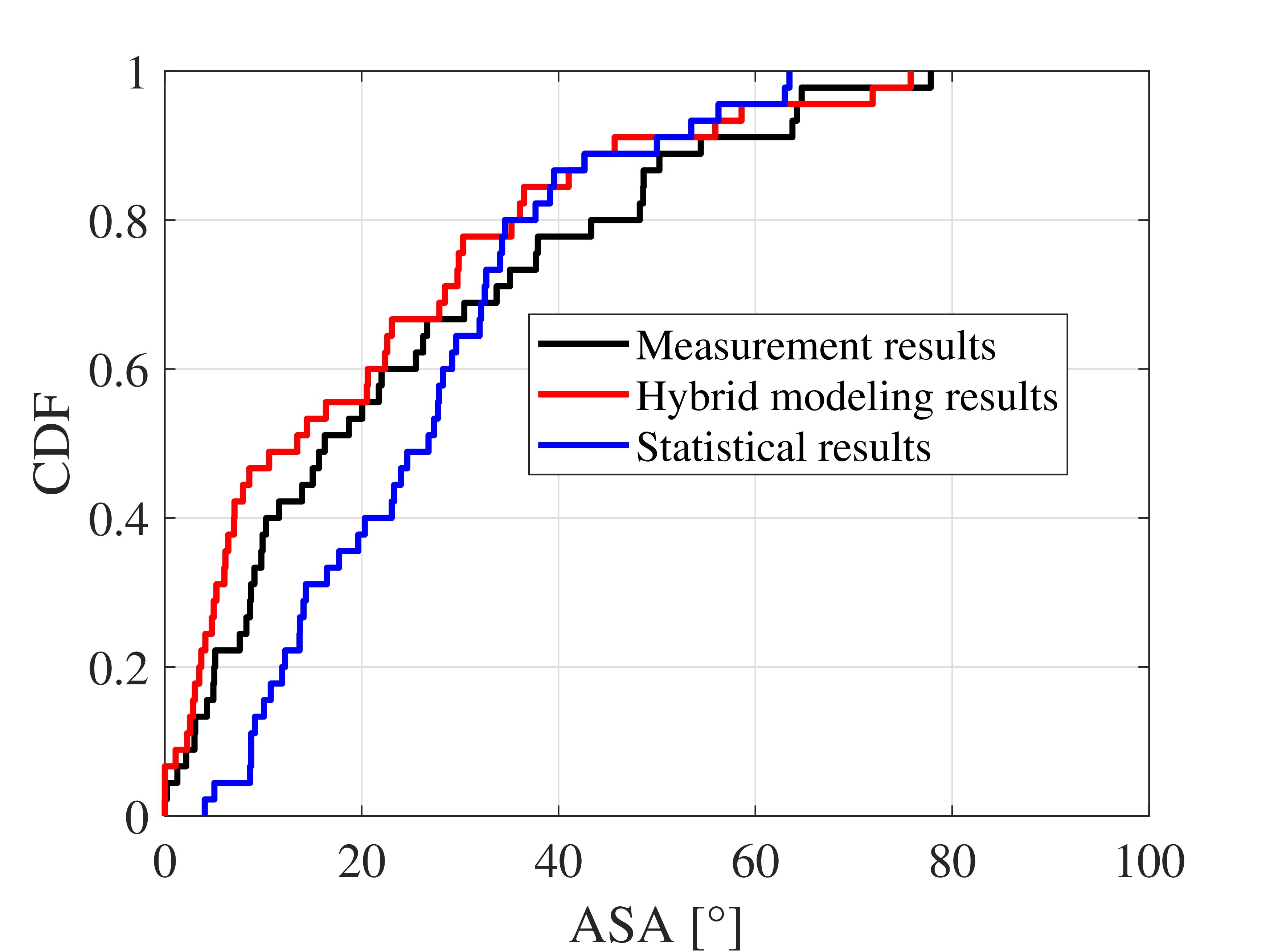}  
    }
    \quad
    \subfloat[NLoS.] {   
    \includegraphics[width=0.8\columnwidth]{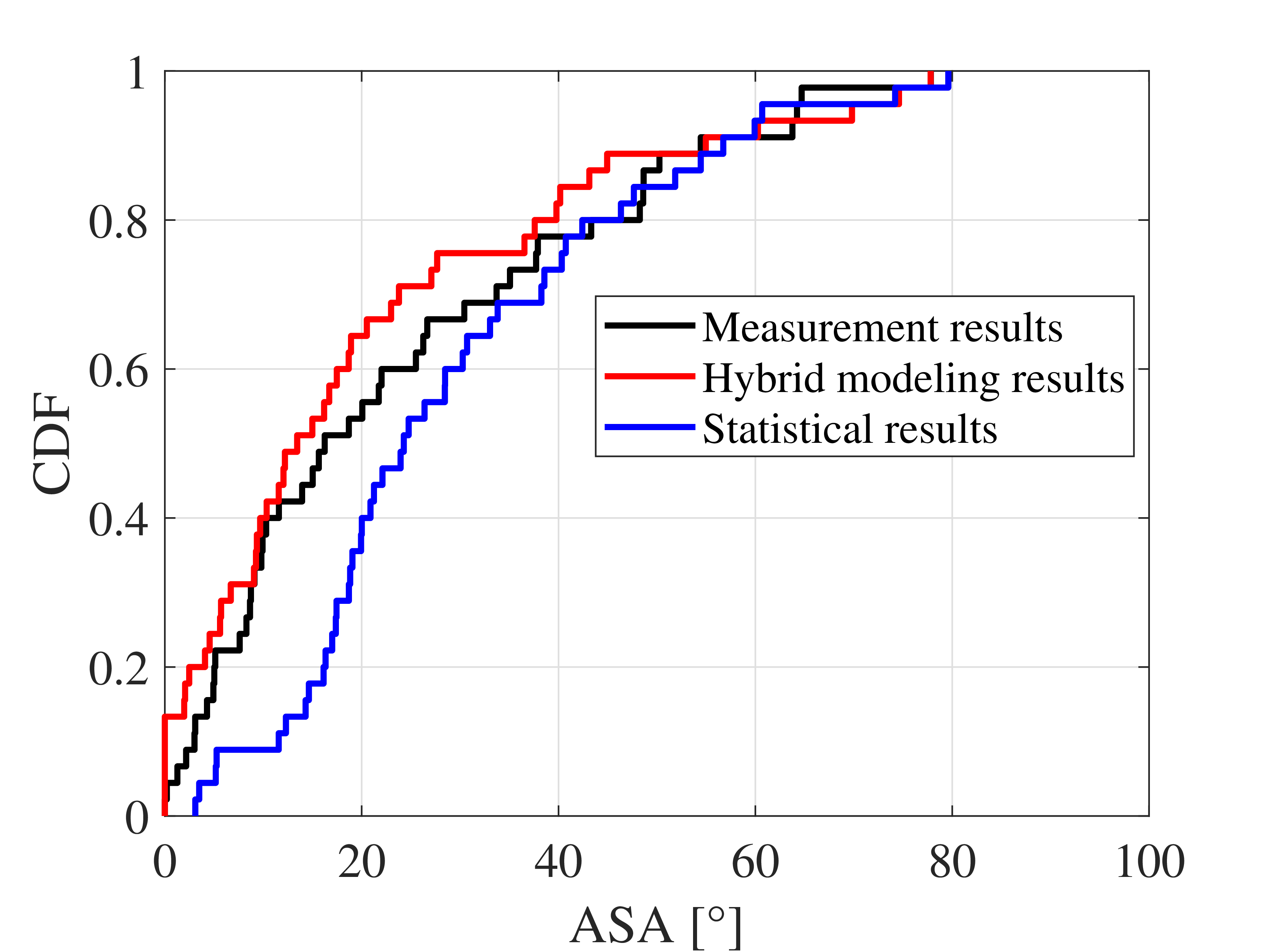}  
    }
    \caption{The ASA results with different channel models.}
    \label{fig:asacomp}
    \vspace{-0.5cm}
\end{figure}
\par To validate the effectiveness of the proposed DTECM, channel realizations are re-generated at the same Tx-Rx locations where the channel measurements were conducted. In addition, the pure statistical model implemented using QuaDriGa is adopted as a benchmark for comparison, with its configuration parameters listed in Table~\ref{tab:fitp}. To assess model performance, the delay and angular spreads are calculated and illustrated in Figs.~\ref{fig:dscomp} and~\ref{fig:asacomp}. As observed, both the DTECM and the statistical model yield delay and angular spread values that closely align with the measurement results. Specifically, in the LoS case, the average delay spread values are \SI{36.73}{ns} for the measurements, \SI{30.80}{ns} for the DTECM, and \SI{32.64}{ns} for the statistical model. In the NLoS case, the respective delay spread values are \SI{48.08}{ns}, \SI{46.68}{ns}, and \SI{49.73}{ns}. Regarding the ASA results, the average values in the LoS scenario are $23.87^\circ$ for the measurements, $20.77^\circ$ for the DTECM, and $29.58^\circ$ for the statistical model. In the NLoS case, the corresponding ASA values are $31.12^\circ$, $21.24^\circ$, and $28.54^\circ$, respectively.
\begin{figure*}[!tbp]
    \centering
    \subfloat[Rx route.]{
    \centering
    \includegraphics[width=0.73\columnwidth]{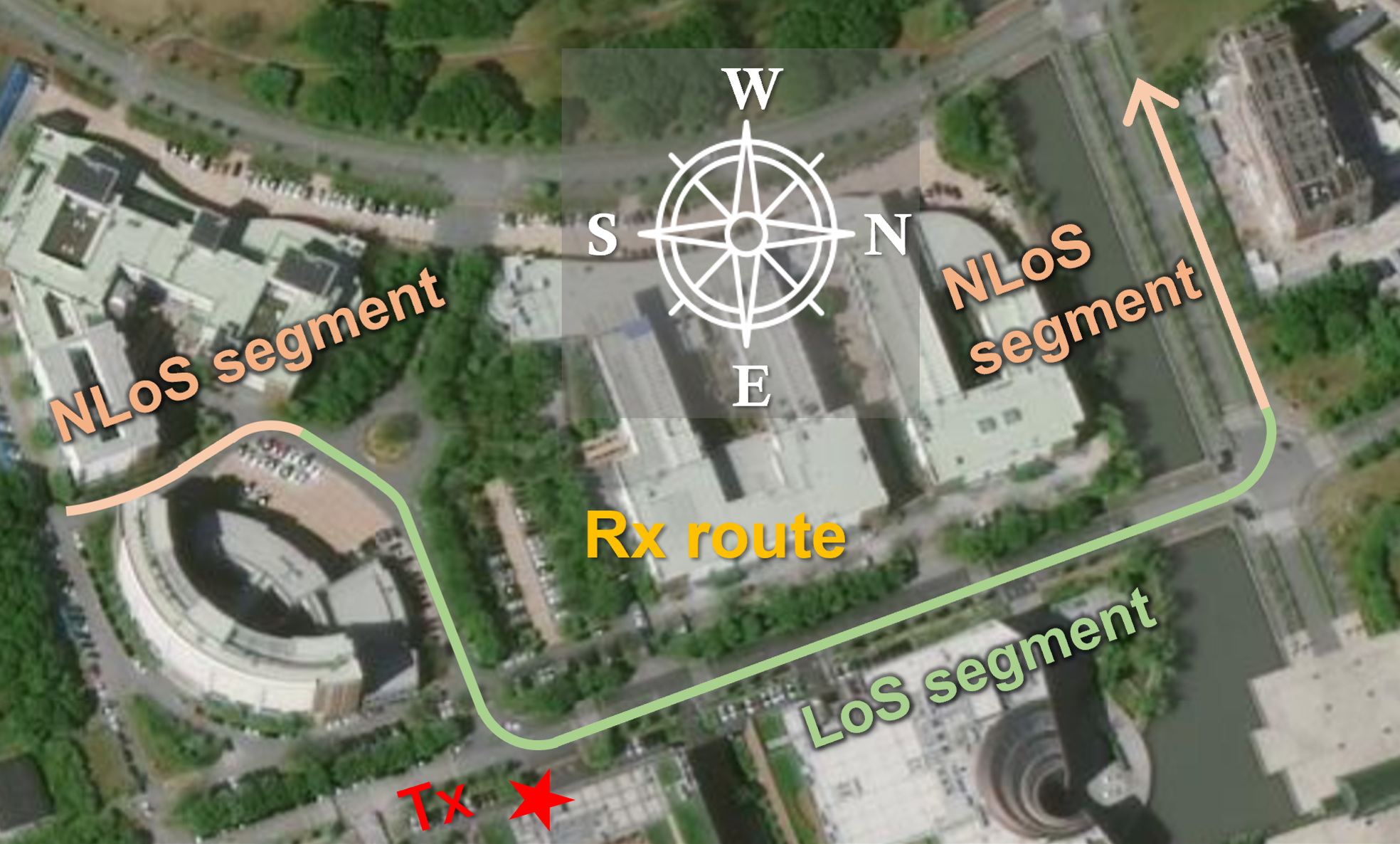}  
    }
    \subfloat[Path loss.]{ 
    \centering
    \includegraphics[width=1.2\columnwidth]{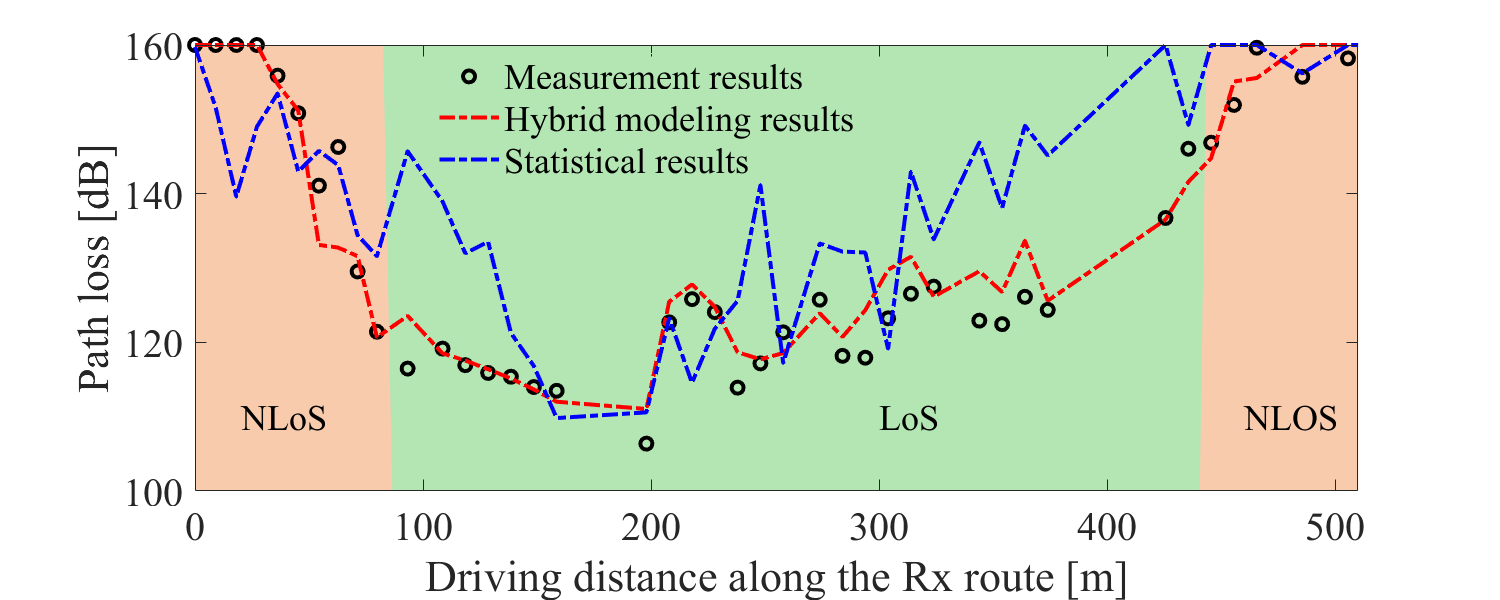}  
    }
    \caption{The path loss modeling results along the investigated Rx route.}
    \label{fig:routepl}
    \vspace{-0.5cm}
\end{figure*}
\par Apart from the statistical distribution of delay and angular spreads, the modeling accuracy of the path loss is also of great significance for effectively characterizing the THz UMa channel. In this regard, an Rx route comprising 42 measurement locations is selected for model performance validation, as illustrated in Fig.~\ref{fig:routepl}(a). Along this Rx route, the LoS condition transitions from an NLoS case at the beginning, to a LoS case in the middle segment, and returns to an NLoS case toward the end. The total length of the route exceeds \SI{500}{m}. The path loss results obtained from the DTECM and the statistical model are presented in Fig.~\ref{fig:routepl}(b), where several key observations can be drawn.
\par First, the DTECM accurately captures the outage locations (PL $>$ \SI{160}{dB}) in the NLoS case, such as the initial \SI{50}{m} segment of the route. This capability is attributed to the ray tracing component of the DTECM. In contrast, the statistical model fails to reproduce this phenomenon, as it generates channel realizations stochastically without incorporating site specific information. Quantitatively, the RMSE between the measurement results and the modeling results in the NLoS case is \SI{4.50}{dB} for the hybrid modeling method, compared to \SI{13.9}{dB} for the statistical modeling method, thereby demonstrating the effectiveness of the proposed DTECM. Second, in the LoS case corresponding to the middle portion of the route, the DTECM significantly outperforms the statistical method in modeling path loss. The statistical method incorporates foliage loss only within the path loss model through a linear function of the Tx Rx distance and a shadow fading component. However, this approach provides only a statistical fit to the measured channel data, resulting in substantial deviations in individual realizations. By contrast, the DTECM derives path loss values from deterministic elements, namely the LoS path obtained via ray tracing and the foliage loss estimated using the digital twin of foliage, leading to much higher modeling accuracy. In particular, the RMSE between the measured and modeled path loss in the LoS case is \SI{3.75}{dB} for the hybrid model and \SI{14.83}{dB} for the statistical model, confirming a substantial reduction in modeling error and validating the superiority of the proposed approach.
\subsection{Link Performance Evaluation in THz UMa}
\begin{figure}
    \centering
    \subfloat[Capacity.] {   
    \includegraphics[width=0.8\columnwidth]{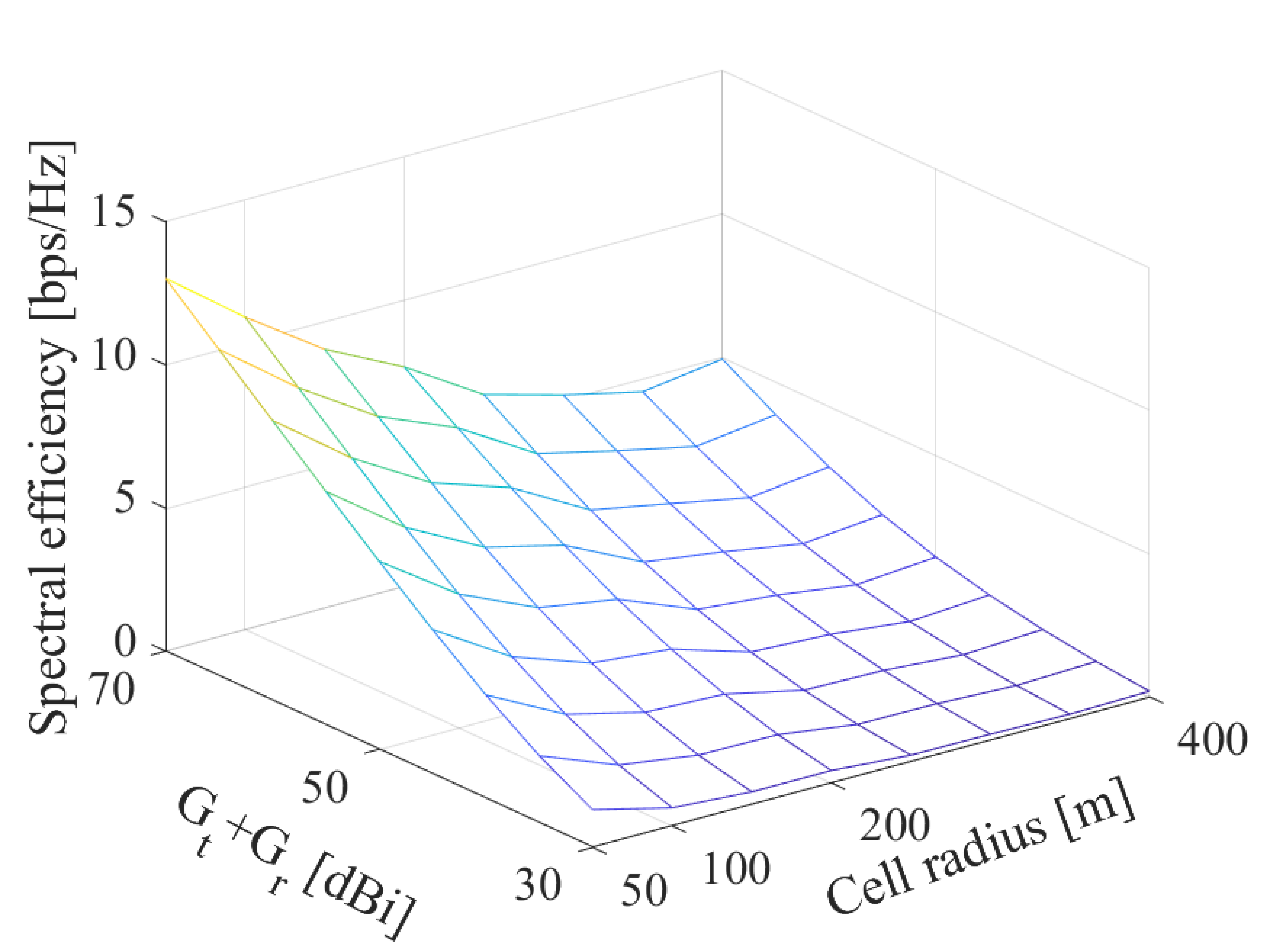}  
    }
    \quad
    \subfloat[Coverage ratio.] {
    \includegraphics[width=0.8\columnwidth]{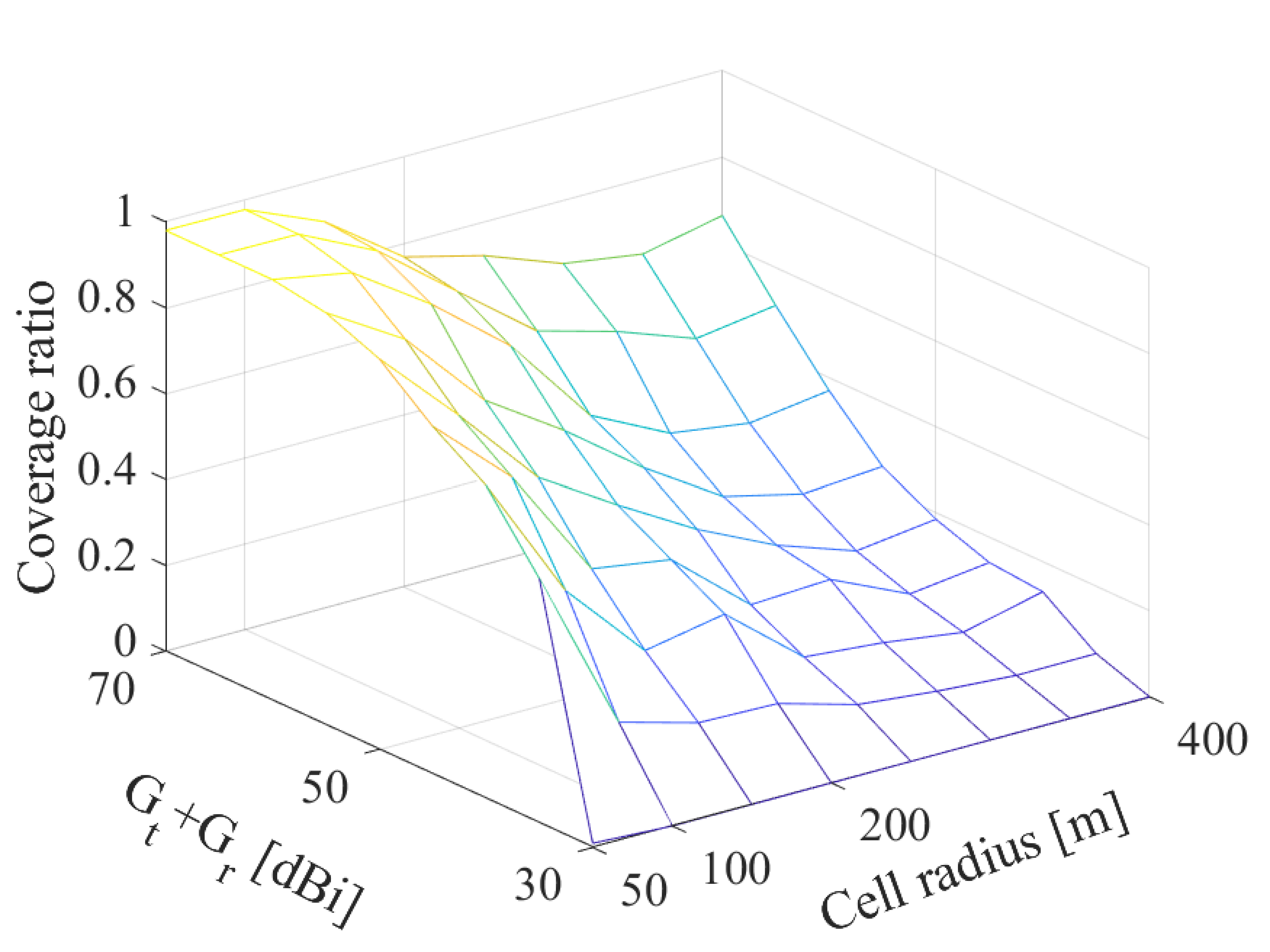}  
    }
    \caption{The link performance evaluation results in the THz UMa based on the DTECM.}
    \label{fig:linkperform}
    \vspace{-0.5cm}
\end{figure}
\par In this section, the link performances in the THz UMa scenario, including the SE and coverage ratio, are evaluated based on the DTECM. Specifically, given a predefined cell radius, the Tx position is set to match the location used during the channel measurement campaign, while the Rx positions are randomly distributed within the defined circular cell area. It is important to note that only the half-circle region in front of the Longbin Building is considered, as the Tx antenna is positioned below the height of the building and thus cannot provide coverage to the area behind it. With these randomly generated Rx positions, the DTECM is employed to generate the corresponding channel realizations and compute the associated path loss. Consequently, the SNR received at a particular Rx location is calculated as follows,
\begin{equation}
    \gamma=\frac{P_tG_tG_r 10^{-\text{PL}/10}}{FkTB},
\end{equation}
where $P_t$ denotes the transmitter power. $G_t$ and $G_r$ stand for the antenna gains of the Tx and Rx. Moreover, $\text{PL}$ is the path loss value in dB that is modeled by using the DTECM. $F$ is the noise figure of the receiver. $k$ stands for the Boltzmann constant and $T$ and $B$ are the temperature and bandwidth.
\par Specifically, the parameters of practical THz communication links are considered as follows. The transmit power is set to \SI{13}{dBm}, while the system temperature and bandwidth are assumed to be \SI{300}{K} and \SI{2}{GHz}, respectively. The noise figure is configured as \SI{10}{dB}. Regarding antenna gains, the total gain—defined as the summation of the Tx and Rx antenna gains—is treated as a variable, since the gain values can vary significantly depending on the types and apertures of the antennas employed at both ends.
\par The average SE and coverage ratio are adopted as the key performance metrics, which are computed as follows,
\begin{align}
        \overline{\text{SE}}&=\frac{\sum_{n_{\text{Rx}}}^{N_{\text{Rx}}}B\log_2(1+\gamma_{n_{\text{Rx}}})}{N_{\text{Rx}}B},\\
        r_\text{cov}&=\text{Pr}(\gamma_{n_{\text{Rx}}}>\gamma_{\text{th}}),
\end{align}
where $\gamma_{n_{\text{Rx}}}$ denotes the received SNR at the $(n_{\text{Rx}})^\text{th}$ randomly generated Rx location, and $N_{\text{Rx}}$ represents the total number of Rx locations considered. Moreover, $\text{Pr}(\cdot)$ denotes the probability of a given event. It is worth noting that only single-input-single-output (SISO) systems are considered in this preliminary performance evaluation. The actual communication performance in THz UMa scenarios could be further enhanced by incorporating MIMO techniques.
\par Based on Monte Carlo simulations, the average SE and coverage ratio results in the THz UMa scenario are illustrated in Fig.~\ref{fig:linkperform}, from which several key observations can be made. First, with high antenna gains and a small cell radius, e.g., a total antenna gain of \SI{70}{dB} and a cell radius of \SI{50}{m}, the average SE exceeds \SI{10}{bps/Hz}, demonstrating the strong potential of the THz band for supporting ultra-high-speed communications. Second, when the antenna gains are reduced and the cell size is increased, the SE experiences a significant degradation. For example, with \SI{30}{dB} antenna gains, the average SE drops to as low as \SI{1.32}{bps/Hz} for a \SI{50}{m} cell radius and \SI{0.19}{bps/Hz} for a \SI{400}{m} cell radius. This highlights the critical importance of employing high-gain antenna techniques to enable high data rate transmissions in THz UMa scenarios. Third, from the perspective of coverage ratio, both antenna gains and cell radius play a crucial role in determining the coverage performance. On one hand, increasing the antenna gains leads to a significant improvement in coverage ratio, owing to the enhanced received SNR. On the other hand, as discussed in Sec.~\ref{sec:scandpa}, many NLoS locations exhibit received signal power levels below the considered path gain threshold of \SI{160}{dB} (corresponding to \SI{-6}{dB} SNR with \SI{70}{dBi} antenna gains). Consequently, effective coverage enhancement techniques are essential to ensure signal reliability in these NLoS regions. Promising approaches include the use of intelligent reflecting surfaces (IRS)\cite{Su2023Wideband,Xue2024Survey} and non-intelligent reflecting surfaces (NIRS)\cite{ju2023ghz,Han2024Still}.
\section{Conclusion}
\label{sec:conclude}
\par In this paper, we conducted extensive channel measurements in the THz UMa scenario at \SI{220}{GHz} using a correlation-based channel sounder. A total of 72 Rx positions were measured, covering LoS, OLoS, and NLoS cases, with the Tx-Rx separation distance reaching up to \SI{410}{m}. Based on the measurement results, we performed a detailed propagation analysis and channel characterization in the THz UMa. The results show that THz wave propagation is significantly influenced by various objects, including buildings, cars, foliage, and metal pillars. Statistical modeling of the channel characteristics reveals that THz channels in the UMa are highly sparse, with weak multipath effects. Furthermore, we created a digital twin of foliage in the angular space based on panoramic images. A KNN-based foliage identification method is proposed to extract foliage from the panoramic images, and the precise foliage positions are determined using camera pose correction. Lastly, we introduced a DTECM, which combines ray-tracing techniques, computer vision methods, and statistical models, outperforming traditional statistical channel models in terms of accuracy. Specifically, the RMSE of path loss is reduced from \SI{14}{dB} using traditional statistical models to approximately \SI{4}{dB} using the DTECM. Link performance evaluation results with the DTECM indicate that high-gain antennas and coverage extension techniques are essential for establishing effective communication links in the THz UMa.
\bibliographystyle{IEEEtran}
\bibliography{IEEEabrv,main}

\end{document}